\documentclass[11pt]{article}

\usepackage[letterpaper, margin=1in]{geometry}

\usepackage{graphicx}
\usepackage{adjustbox}
\usepackage{amsmath}
\usepackage{amsthm}
\usepackage{amssymb}
\usepackage{amsfonts}
\usepackage{mathtools}
\usepackage{url}
\usepackage{float}
\usepackage{multirow}
\usepackage{authblk}
\usepackage{listings}
\usepackage{bbold}
\usepackage[utf8]{inputenc}
\usepackage[english]{babel}
\usepackage{natbib}
\usepackage{apalike}

\newcommand\independent{\protect\mathpalette{\protect\independenT}{\perp}}
\def\independenT#1#2{\mathrel{\rlap{$#1#2$}\mkern2mu{#1#2}}}

\DeclarePairedDelimiterX{\divx}[2]{(}{)}{%
  #1\;\delimsize\|\;#2%
}

\DeclarePairedDelimiterX{\distx}[2]{[}{]}{%
  #1\;\delimsize\|\;#2%
}

\newcommand{\infdiv}{D_f\divx}
\newcommand{\infdivb}{D_f\distx}

\DeclareMathOperator*{\argmax}{arg\,max}
\DeclareMathOperator*{\argmin}{arg\,min}

\providecommand{\keywords}[1]{\textbf{\textit{Keywords --}} #1}

\newtheorem{lemma}{Lemma}
\newtheorem{prop}{Proposition}
\newtheorem{theorem}{Theorem}
\newtheorem{corollary}{Corollary}
\newtheorem{remark}{Remark}

\makeatletter
\g@addto@macro\th@plain{\thm@headpunct{:}}
\makeatother
\theoremstyle{plain}
\newtheorem{assumption}{Assumption}

\begin{document}
\title{A Framework for Covariate Balance using Bregman Distances}
\author{Kevin P. Josey}
\author{Elizabeth Juarez-Colunga}
\author{Fan Yang}
\author{Debashis Ghosh}
\affil{Department of Biostatistics and Informatics, Colorado School of Public Health, University of Colorado Anschutz Medical Campus}

\maketitle

\begin{abstract}
A common goal in observational research is to estimate marginal causal effects in the presence of confounding variables. One solution to this problem is to use the covariate distribution to weight the outcomes such that the data appear randomized. The propensity score is a natural quantity that arises in this setting. Propensity score weights have desirable asymptotic properties, but they often fail to adequately balance covariate data in finite samples. Empirical covariate balancing methods pose as an appealing alternative by exactly balancing the sample moments of the covariate distribution. With this objective in mind, we propose a framework for estimating balancing weights by solving a constrained convex program where the criterion function to be optimized is a Bregman distance. We then show that the different distances in this class render identical weights to those of other covariate balancing methods. A series of numerical studies are presented to demonstrate these similarities.
\end{abstract}


\section{Introduction}\label{intro}

Causal inference has been a topic of intense interest in the statistical literature. The focus of causal inference methodology deals with the issue of how to properly evaluate treatment effects in a non-randomized setting. In many medical and scientific studies, randomization cannot be performed due to logistic, economic, and/or ethical limitations. Under these circumstances, the emergent challenge is the consistent evaluation of treatment effects in the presence of confounding. For example, in \cite{bhagat_national_2017}, a cohort of patients undergoing lung resection surgery are examined to compare the rate of unplanned readmission following thoracoscopic versus open anatomic lung resections. The mode of surgery is not randomized and there are several pre-operative characteristics which inform the method of surgery, that in turn affect the readmission rate. Methods for causal inference seek to minimize any bias induced by these confounding variables.

Two important concepts in causal inference are the potential outcomes model \citep{splawa-neyman_application_1990, rubin_estimating_1974} and the propensity score \citep{rosenbaum_central_1983}. The potential outcomes approach provides a powerful tool for conceptualizing, estimating, and performing inference regarding causal effects. An overview for implementing the potential outcomes model can be found in \cite{imbens_causal_2015}. They demonstrate that a natural quantity which regularly arises when balancing potential confounders between experimental groups in observational studies is the propensity score \citep{rosenbaum_central_1983}. The propensity score is defined as the probability of receiving treatment given a set of measured covariates. Based on the assumptions underlying the potential outcomes model and the propensity score, causal inference proceeds in the following stages: (a) a propensity score model is fit using the observed data; (b) diagnostics for covariate balance using the propensity score are evaluated; and (c) estimates of the causal effect are produced by conditioning on the propensity score. Iterating between steps (a) and (b) is often necessary to ensure the homogeneity of the propensity score adjusted covariate distributions.

A key goal for the propensity score model is to achieve covariate balance, which means that the distribution of confounders between the treated and control groups are equivalent. From \cite{rosenbaum_central_1983}, the assumptions of strongly ignorable treatment assignment (defined in Section \ref{potential}), in conjunction with the definition of the propensity score, imply that adjustment on the propensity score alone will theoretically achieve balance. However, this result is based on the population propensity score and does not necessarily hold in finite samples. There have been numerous approaches that address the issue of balancing empirical covariate distributions using weighting estimators. We refer to the weights produced by these methods as \textit{balancing weights}. One popular method is to construct propensity scores with covariate balance built into the estimation procedure. \cite{imai_covariate_2013} and \cite{fan_improving_2018} introduced the covariate balance propensity score (CBPS) and its subsequent improvement (iCBPS), both of which use generalized methods of moments to fit a logit model with covariate balance serving as an auxiliary condition. Any resulting estimate of the propensity score will automatically achieve balance by construction.

In the political science literature, \cite{hainmueller_entropy_2012} uses maximum entropy density estimation to find balancing weights to estimate the average treatment effect of the treated. The algorithm, termed entropy balancing, finds the vector of balancing weights that minimize the normalized relative entropy from a vector of sampling weights subject to a set of linear equality constraints about the moments of the covariate distribution. Recent work by \cite{zhao_entropy_2017} shows how this algorithm enjoys a double-robustness property. The general idea of double-robust estimation is to combine covariate information about the treatment assignment and the outcome model into the weighting estimator \citep{bang_doubly_2005, kang_demystifying_2007}. If at least one model is correctly specified, then the resulting causal effect estimate is consistent. When both the outcome and treatment models are correctly specified, then the estimate achieves the semiparametric efficiency bound described by \cite{hahn_role_1998}. Entropy balancing is limited to finding balancing weights to estimate the average treatment effect of the treated, leaving gaps in the procedure for developing doubly-robust estimators of other estimands. This issue is related to the choice of \cite{hainmueller_entropy_2012} to optimize the normalized relative entropy. By changing the criterion distance function in a convex optimization problem, similar to the one presented by \cite{hainmueller_entropy_2012}, we can draw parallels to other covariate balancing methods. A similar idea is proposed by \cite{zhao_covariate_2019}, who shows that CBPS and entropy balancing can be generalized by modifying the score function derived from the respective covariate balance problem. Calibration estimators \citep{deville_calibration_1992} also produce balancing weights using constrained convex optimization techniques. The proposed methods in \cite{ chan_globally_2015} implicitly extends entropy balancing to include other distance functions. However, they restrict their attention to a nonparametric setting, characterizing their methodology as a departure from the propensity score literature.

Our aim is to extend the work of \cite{hainmueller_entropy_2012}, \cite{imai_covariate_2013}, and \cite{fan_improving_2018} for finding balancing weights that facilitate causal effect estimation when the treatment assignment is not determined by a logit model. We do so by demonstrating how balancing weights can be computed from Bregman distances \citep{bregman_relaxation_1967}. Bregman distances have multiple geometric properties that allow for easy estimation of the balancing weights. This geometric interpretation of balancing weights complements the implicit geometry found in classic semiparametric inference. Using the results of our framework, we prove that CBPS \citep{imai_covariate_2013} and iCBPS \citep{fan_improving_2018} are doubly-robust estimators of the average treatment effect while assuming the propensity scores follow a logit model. As an extension to CBPS, we propose an estimator for balancing weights akin to the overlap weights discussed by \cite{li_balancing_2018}. We also show how our framework is consistent with the calibration estimator approach of \cite{chan_globally_2015}, thereby bridging the empirical covariate balancing methods of entropy balancing, CBPS, iCBPS, and calibration estimators. We are interested in these methods in particular as they do not incorporate a model of the outcome process into their designs in the spirit of \cite{rubin_for_2008}.

The outline of this article is as follows. Section \ref{notation} defines the general notation and assumptions that will be applied throughout the manuscript. Section \ref{bregman} describes the methods for finding balancing weights by solving a constrained optimization problem using Bregman distances as the criterion function. Section \ref{related} describes the similarities between our method and other covariate balancing methods. Section \ref{numerical} summarizes results from two simulation studies comparing different covariate balancing methods. This section also contains the results for a replication study of \cite{bhagat_national_2017} using a variety of different covariate balancing methods. The real data set illustrates the importance of selecting appropriate covariate balancing methods. Finally, Section \ref{discussion} concludes with a discussion of the framework and future work.

\section{Background and Preliminaries}\label{notation}

\subsection{Notation and Definitions}

Parameters will be denoted using Greek letters, whereas random variables will be denoted with Roman letters. Boldface letters will denote vectors and matrices while non-boldface letters represent scalars. For a matrix $\mathbf{A}$ the transpose is written as $\mathbf{A}^{T}$. The symbol $\nabla f$ denotes the gradient of a function $f$. Let $\mathbf{1}_n$ denote the $(n \times 1)$ vector with each entry equal to one. Similarly, let $\mathbf{0}_n$ denote the $(n \times 1)$ vector with each entry equal to zero.

Let $\mathbf{X}$ denote a vector of real-valued covariate measurements, $Z$ denote the random treatment assignment with support $\{0,1\}$, and $Y$ denote the real-valued outcome variable. The independent sampling units will be indexed by $i = 1,2,\ldots,n$. The $(n \times 1)$ vector of balancing weights will be written as $\mathbf{p} \equiv (p_1, p_2, \ldots, p_n)^{T}$ while the $(n \times 1)$ vector of sampling weights will be written as $\mathbf{q} \equiv (q_1, q_2, \ldots, q_n)^{T}$. We will often write $p_i = p(\mathbf{X}_i)$ for the $i$th subject to emphasize the fact that the balancing weights are conditioned on the covariates. Define $\{c_j(\mathbf{X}); j = 1,2,\ldots, m\}$, as a set of functions that generate linearly independent features to be balanced between treatment groups. We will refer to these quantities as \textit{balance functions}.

\subsection{Potential Outcomes Model}\label{potential}

Potential outcomes provide a convenient framework for conceptualizing causal effects. This framework was first introduced by \cite{splawa-neyman_application_1990} for randomized experiments. The concepts and assumptions necessary to extend this framework to observational data were later formalized by \cite{rubin_estimating_1974}. The potential outcomes are denoted with a vector $\left[Y(0), Y(1)\right]^{T}$ with $Y(0)$ and $Y(1)$ corresponding to the counterfactual outcome when $Z = 0$ and $Z = 1$, respectively. The conditional expectations for the potential outcomes are denoted with $\mu_0(\mathbf{X}) \equiv \mathbb{E}[Y(0)|\mathbf{X}]$ and $\mu_1(\mathbf{X}) \equiv \mathbb{E}[Y(1)|\mathbf{X}]$. The random outcome is defined by the transformation $Y \equiv Z Y(1) + (1-Z)Y(0)$. Some common causal estimands are the population average treatment effect (ATE), $\tau_{\text{ATE}} \equiv \mathbb{E}\left[Y(1) - Y(0)\right]$, and the population average treatment effect of the treated (ATT),  $\tau_{\text{ATT}} \equiv \mathbb{E}\left[Y(1) - Y(0)| Z = 1\right]$. In any case, the causal effects are non-identifiable as one of the two required potential outcomes is always missing. This simple observation is the fundamental problem of causal inference. We adopt the setting proposed by \cite{rosenbaum_central_1983} who describe a set of assumptions that will allow us to find consistent estimates of the treatment effect in observational studies. This includes the following assumptions about the data.

\begin{assumption}[Strong Ignorability]\label{sita} $[Y(0), Y(1)]^{T} \independent Z|\mathbf{X}$. \end{assumption} The strong ignorability assumption requires that the vector of potential outcomes be independent of the treatment assignment when we condition on the covariates. This assumption further implies that there is no unmeasured confounding. The implication of Assumption \ref{sita} along with the definition of the propensity score as a balance criterion allows us to conclude that $[Y(0), Y(1)]^{T} \independent Z|\pi(\mathbf{X})$, where $\pi(\mathbf{X}) \equiv \Pr\{Z = 1|\mathbf{X}\}$ denotes the propensity score \citep{rosenbaum_central_1983}.

\begin{assumption}[Positivity]\label{positivity} $0 < \Pr\{Z = 1|\mathbf{X}\} < 1$ for all $\mathbf{X}$. \end{assumption} The treatment positivity assumption requires the probability that a subject is assigned to the treatment group as opposed to the control group be bounded away from zero and one. Since $\Pr\{Z = 1|\mathbf{X}\}$ must be estimated from the covariates, then Assumption \ref{positivity} equivalently amounts to requiring sufficient overlap between the covariate distributions of the two treatment groups. The feasibility of the convex optimization problems that we will introduce later on are deeply intertwined with Assumption \ref{positivity}. Without sufficient overlap, the estimated balancing weights will either not exist, or be unstable and produce biased estimates of the causal effect. 

\subsection{Horvitz-Thompson Estimator}\label{HT-estimator}

The Horvitz-Thompson class of estimators \citep{horvitz_generalization_1952, hirano_efficient_2003} frequently appears in the causal inference literature. For example, the Horvitz-Thompson estimator for the average treatment effect is
\begin{equation}\label{HT-ATE}
\hat{\tau}_{\text{ATE}} = \frac{1}{n}\sum_{i = 1}^n \left[\frac{Z_iY_i}{\pi(\mathbf{X}_i)} - \frac{(1 - Z_i)Y_i}{1 - \pi(\mathbf{X}_i)}\right]
\end{equation}
while the Horvitz-Thompson estimator for $\tau_{\text{ATT}}$ is
\begin{equation}\label{HT-ATT}
\hat{\tau}_{\text{ATT}} = \frac{1}{n_1}\sum_{i = 1}^n \left[Z_iY_i - \frac{\pi(\mathbf{X}_i)(1 - Z_i)Y_i}{1 - \pi(\mathbf{X}_i)}\right]
\end{equation}
where $n_1 = \sum_{i = 1}^n Z_i$. \cite{hahn_role_1998} was able to show that (\ref{HT-ATE}) and (\ref{HT-ATT}) have optimal asymptotic properties for estimating $\tau_{\text{ATE}}$ and $\tau_{\text{ATT}}$. Even when we substitute a consistent estimator of the propensity score into (\ref{HT-ATE}), the estimator for $\tau_{\text{ATE}}$ remains consistent and achieves the semiparametric efficiency bound.

A more general form for causal effect estimation is
\begin{equation}\label{HT}
\hat{\tau} = \sum_{i = 1}^n \frac{(2Z_i - 1)p(\mathbf{X}_i)Y_i}{\sum_{i = 1}^n p(\mathbf{X}_i)Z_i},
\end{equation} 
which accommodates several different estimands through the choice of $p(\mathbf{X})$. For example, we will see in Section \ref{ATT} that the estimator for $\tau_{\text{ATT}}$ is similar to estimators of $\tau_{\text{ATE}}$ with additional constraints placed on the balancing weights so that $p(\mathbf{X}) = q$ whenever $Z = 1$. Equations (\ref{HT-ATE}) and (\ref{HT-ATT}) provide direction for identifying $p(\mathbf{X})$ within (\ref{HT}) in order to estimate $\tau_{\text{ATE}}$ and $\tau_{\text{ATT}}$, respectively. If the propensity score is known, $\tau_{\text{ATE}}$ can be estimated by setting $p(\mathbf{X}) = \pi(\mathbf{X})^{-1}$ when $Z = 1$ and $p(\mathbf{X}) = [1 - \pi(\mathbf{X})]^{-1}$ when $Z = 0$. We can also find an estimator for $\tau_{\text{ATT}}$ by setting $p(\mathbf{X}) = \pi(\mathbf{X})[1 - \pi(\mathbf{X})]^{-1}$ when $Z = 0$ and $p(\mathbf{X}) = 1$ when $Z = 1$. When the propensity score is unknown, finding an estimator for $p(\mathbf{X})$ that produces consistent estimates of $\tau_{\text{ATE}}$ and $\tau_{\text{ATT}}$ is relatively straightforward. Estimating balancing weights that also preserve the efficiency of $\hat{\tau}_{\text{ATE}}$ and $\hat{\tau}_{\text{ATT}}$ is a more challenging proposition.

\section{Bregman Distances}\label{bregman}

\subsection{Definition}\label{breg-distances}

\begin{figure}[H]
	\centering
	\includegraphics[scale = 0.7]{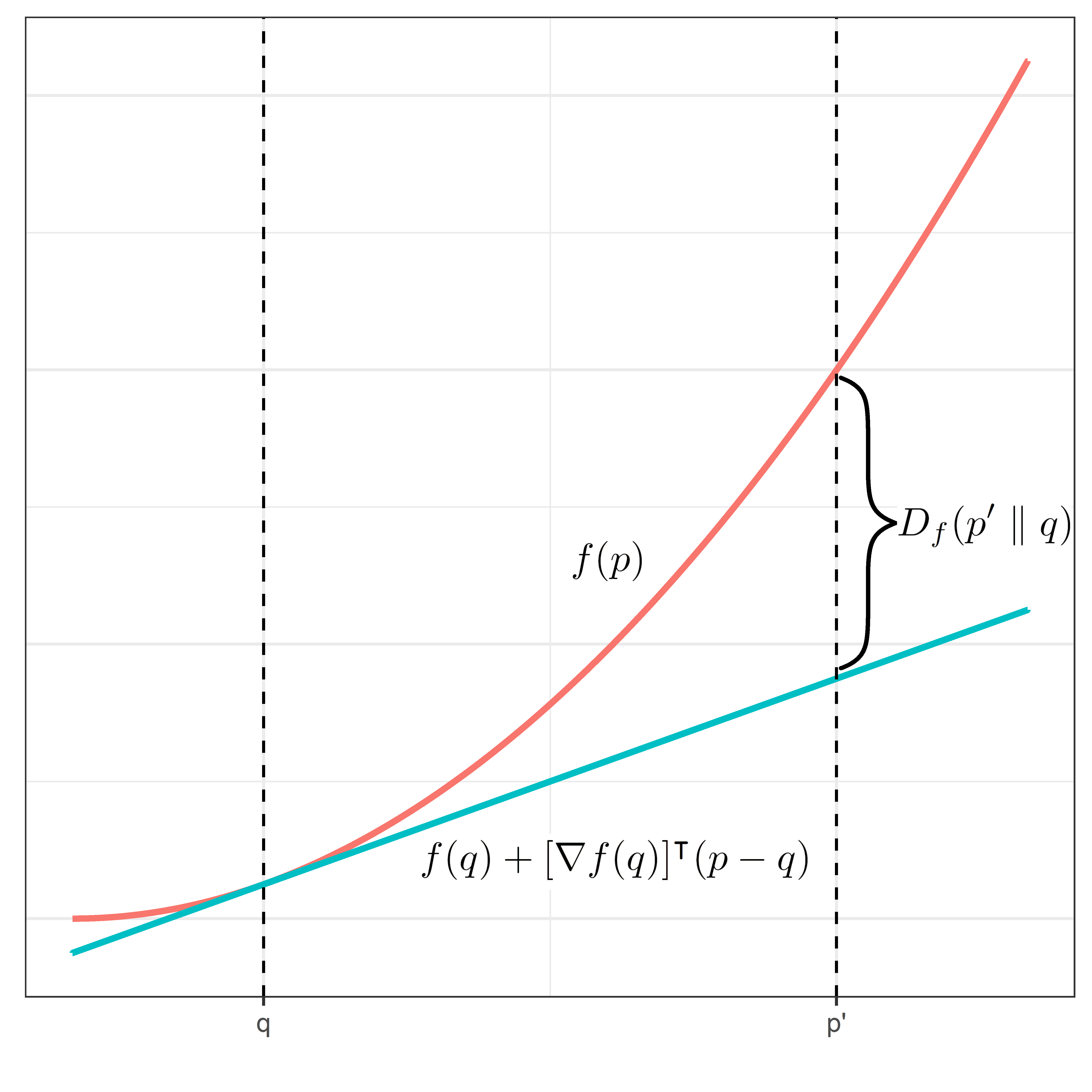}
	\caption{Example of a Bregman distance for one-dimensional $p', q \in \Delta$. The function $f(p)$ (red line) is strictly convex over $p \in \Delta$. The line tangent to $f$ at $q$ is the blue line. The Bregman function is the distance between the red and blue lines at the point $p'$.}\label{breg-plot}
\end{figure}

Let $\Delta^n \subseteq \Re^n$ be a non-empty, convex, and open set with closure $\bar{\Delta}^n$. Define $f: \bar{\Delta}^n \rightarrow \Re$ to be a continuously differentiable, strictly convex function. The Bregman distance generated by the function $f$ is the difference between $f$ evaluated at $\mathbf{p} \in \bar{\Delta}^n$ and the first-order Taylor series approximation of $f$ about $\mathbf{q} \in \Delta^n$, evaluated at $\mathbf{p}$. In other words, a Bregman distance $D_f: \bar{\Delta}^n \times \Delta^n \rightarrow \Re$  may be defined as \[ \infdiv{\mathbf{p}}{\mathbf{q}} \equiv f(\mathbf{p}) - f(\mathbf{q}) - [\nabla f(\mathbf{q})]^{T}  (\mathbf{p} - \mathbf{q}). \] Bregman distances are often used to measure the convexity associated with $f$. Since $f$ is strictly convex over $\bar{\Delta}^n$, it follows that for $\mathbf{p} \in \bar{\Delta}^n$ and $\mathbf{q} \in \Delta^n$, $\infdiv{\mathbf{p}}{\mathbf{q}} \ge 0$ with equality holding if and only if $\mathbf{p} = \mathbf{q}$. This implies that $\infdiv{\mathbf{p}}{\mathbf{q}}$ is also strictly convex. A more complete definition of Bregman distances can be found in Chapter 2 of \cite{censor_parallel_1998}, which includes additional properties that $D_f$ must satisfy which are not mentioned here. A visual representation of a Bregman distance can be found in Figure \ref{breg-plot}.

One of the most common examples of a Bregman distance is the unnormalized relative entropy. Let $f(\mathbf{p}) = \sum_{i = 1}^n p_i \log\left(p_i\right)$ for $\mathbf{p} \in [0,\infty)^n$. We assume $0\log(0) = 0$ so that the domain of $f$ includes the boundary points contained within the closure of $\Delta^n$. The resulting Bregman distance is written as \[ \infdiv{\mathbf{p}}{\mathbf{q}} = \sum_{i = 1}^n \left[p_i\log\left(\frac{p_i}{q_i}\right) - p_i + q_i\right]. \] The Euclidean distance is another example of a Bregman distance. By selecting $f(\mathbf{p}) = \sum_{i = 1}^n p_i^2/2$ for $\mathbf{p} \in \Re^n$ we get \[ \infdiv{\mathbf{p}}{\mathbf{q}}  = \sum_{i = 1}^n \frac{(p_i - q_i)^2}{2}.\]

In order to simplify the presentation of the methods, we will only consider Bregman distances that are separable. This means $\infdiv{\mathbf{p}}{\mathbf{q}} = \sum_{i = 1}^n \infdiv{p_i}{q_i}$. Note that both the unnormalized relative entropy and the Euclidean distance are separable. Since we require positive weights, we also restrict our focus to convex functions where $\Delta^n \subseteq [0,\infty)^n$ in order to avoid setting additional constraints for $\mathbf{p} \ge \mathbf{0}_n$. Notice that the domain of the unnormalized relative entropy satisfies this condition while the domain of the Euclidean distance does not. In addition, we will assume throughout that the sampling weights $\mathbf{q} \in \Delta^n$ are fixed by design and known.

\subsection{Constrained Optimization and Duality}\label{constrain}

For some $\mathbf{q} \in \Delta^n$, the value $\hat{\mathbf{p}} \in \bar{\Delta}^n$ that minimizes $\infdiv{\mathbf{p}}{\mathbf{q}}$ in an unconstrained setting is $\hat{\mathbf{p}} = \mathbf{q}$. In covariate balance problems, we specify a set of linear constraints that the optimal solution must satisfy. Consider the constrained convex optimization problem to
\begin{equation}\label{primal}
\begin{split} 
\text{minimize} &\quad \sum_{i = 1}^n \infdiv{p_i}{q_i} \\ 
\text{subject to} &\quad \mathbf{A}^{T}\mathbf{p} = \mathbf{b}
\end{split}
\end{equation}
where $\mathbf{A}$ is a linearly independent $(n \times m)$ matrix that forms the basis of a linear subspace that defines the constraints of the program and $\mathbf{b}$ is an $(m \times 1)$ vector denoting the margins of those constraints. The entries of $\mathbf{A}$ and $\mathbf{b}$ are denoted with $a_{ij} \in \Re$ and $b_j \in \Re$ ($i = 1,2,\ldots,n$ and $j = 1,2,\ldots,m$), respectively. Equation (\ref{primal}) is often referred to as the primal problem and the corresponding solution is referred to as the primal solution. We denote the set of feasible primal solutions that satisfy the linear constraints in (\ref{primal}) as $\Omega \equiv \left\{\mathbf{p} : \mathbf{A}^{T}\mathbf{p} = \mathbf{b}\right\}$. 

Geometrically, the solution to the primal problem is the point
\begin{equation}\label{gen-proj}
\hat{\mathbf{p}} \equiv \argmin_{\mathbf{p} \in \Omega \cap \bar{\Delta}^n} \infdiv{\mathbf{p}}{\mathbf{q}},
\end{equation} 
which is the generalized projection of $\mathbf{q} \in \Delta^n$ into $\Omega$. Note that $\Omega \cap \bar{\Delta}^n$ is sometimes empty. One solution to avoid this issue is to choose $\mathbf{b} = \mathbf{A}^{T}\tilde{\mathbf{p}}$ where $\tilde{\mathbf{p}} \in \bar{\Delta}^n$ so that $\tilde{\mathbf{p}} \in \Omega \cap \bar{\Delta}^n$. For example when using the unnormalized relative entropy, we might set $\mathbf{b} = \mathbf{0}_m$ with $\tilde{\mathbf{p}} = \mathbf{0}_n$. However, this condition is not so obvious for some of the covariate balancing problems that we will encounter later on. Instead, we will assume $\Omega \cap \bar{\Delta}^n \ne \emptyset$ throughout. Lemma \ref{unique-thm} proves that the generalized projection is unique. The proof appears in Section 2.1 of \cite{censor_parallel_1998}. For the sake of completeness, a version of this proof is also found in the online supplement.

\begin{lemma}\label{unique-thm}
Suppose $\Omega \cap \bar{\Delta}^n \ne \emptyset$. Then the generalized projection of $\mathbf{q}$ into $\Omega$, defined in (\ref{gen-proj}), is unique.
\end{lemma}

When $\hat{\mathbf{p}} \in \Omega \cap \Delta^n$, the primal problem can be solved by introducing a vector of Lagrangian multipliers. With Lagrangian multipliers, we can formulate the Lagrangian $L: \bar{\Delta}^n \times \Delta^n \times \Re^m \rightarrow \Re$ for any constrained optimization problem in the form of (\ref{primal}) as 
\begin{equation}\label{lagrange}
L(\mathbf{p}, \mathbf{q}, \boldsymbol{\lambda}) \equiv \infdiv{\mathbf{p}}{\mathbf{q}} + (\mathbf{A}^{T}\mathbf{p} - \mathbf{b})^{T}\boldsymbol{\lambda}.
\end{equation} 
Optimizing the Lagrangian with respect to $\mathbf{p} \in \Delta^n$ and $\boldsymbol{\lambda} \in \Re^m$ is an unconstrained problem equivalent to (\ref{primal}). Instead of finding the point $\hat{\mathbf{p}} \in \Omega \cap \Delta^n$ that minimizes $\infdiv{\mathbf{p}}{\mathbf{q}}$, we find the vector $\hat{\mathbf{p}}$ that minimizes the Lagrangian with respect to $\mathbf{p} \in \Delta^n$ and $\hat{\boldsymbol{\lambda}}$ that maximizes the Lagrangian with respect to $\boldsymbol{\lambda} \in \Re^m$. In other words, the optimal solution forms a saddle point on (\ref{lagrange}) over the space $\Delta^n \times \Re^m$ for a fixed $\mathbf{q} \in \Delta^n$.

The following Propositions are used in tandem to obtain balancing weights for treatment effect estimation. A necessary condition for these Propositions is that the function $f$ which generates the Bregman distance $D_f$ be zone consistent with respect to $\Omega$. This means that for any $\mathbf{q} \in \Delta^n$, the Bregman distance produced by $f$ has its generalized projection of $\mathbf{q}$ into $\Omega$ contained within the open set $\Delta^n$.

\begin{prop}\label{global-min} 
Assume that $f$ is zone consistent with respect to $\Omega$. Let $\hat{\mathbf{p}} \in \Omega \cap \Delta^n$ be the generalized projection of $\mathbf{q}$ into $\Omega$, as defined in (\ref{gen-proj}). Then $\hat{\mathbf{p}}$ is uniquely determined by  \[ P_f\left(\mathbf{q}, \mathbf{A}\hat{\boldsymbol{\lambda}}\right) \equiv (\nabla f)^{-1}\left(\nabla f(\mathbf{q}) - \mathbf{A}\hat{\boldsymbol{\lambda}}\right) \] where $\hat{\boldsymbol{\lambda}} \in \Re^m$ is also unique.
\end{prop} 

The proof of Proposition \ref{global-min} can be found in the online supplement. Given the result of Proposition \ref{global-min}, the convex optimization problem can be solved by estimating $\hat{\boldsymbol{\lambda}} \in \Re^m$ with the dual problem, which is to
\begin{equation}\label{dual}
\begin{split}
\text{maximize} &\quad L\left[P_f(\mathbf{q}, \mathbf{A}\boldsymbol{\lambda}), \mathbf{q}, \boldsymbol{\lambda}\right] \\
\text{subject to} &\quad \boldsymbol{\lambda} \in \Re^m.
\end{split}
\end{equation} 
Propositions \ref{global-min} and \ref{dual-thm} imply that the primal solution to (\ref{primal}) can be obtained by plugging the solution to the dual problem into the generalized projection. The proof of Proposition \ref{dual-thm} is adapted from Section 3.4 of \cite{bertsekas_nonlinear_1999} and appears in the online supplement.

\begin{prop}\label{dual-thm}
Assume $f$ is zone consistent with respect to $\Omega$. If the primal problem (\ref{primal}) has an optimal solution, then the dual problem (\ref{dual}) also has an optimal solution and the two optimal values are equal.
\end{prop}

\subsection{Balancing Weights for the ATE}\label{ATE}

In this section we outline the general strategy and guidelines for obtaining balancing weights for estimating $\tau_{\text{ATE}}$. The proposed method requires solving the primal problem to
\begin{equation}\label{primal-ate} 
\begin{split} 
\text{minimize} &\quad \sum_{i = 1}^n \infdiv{p_i}{q_i}  \\
\text{subject to} &\quad \sum_{i = 1}^n p_i(2Z_i - 1)c_j(\mathbf{X}_i) = 0 \enskip \text{and} \\
&\quad \sum_{i = 1}^n p_i Z_i c_j(\mathbf{X}_i) = \sum_{i = 1}^n q_ic_j(\mathbf{X}_i) \enskip \text{for all} \enskip j = 1,2,\ldots,m.
\end{split}
\end{equation} 
As mentioned in the previous section, (\ref{primal-ate}) can be solved by optimizing the corresponding Lagrangian, which is defined as
\begin{equation}\label{lagrange-ate} 
\begin{split}
L_{\text{ATE}}(\mathbf{p}, \mathbf{q}, \boldsymbol{\lambda}) &\equiv \sum_{i = 1}^n \infdiv{p_i}{q_i} + \sum_{ j = 1 }^m \lambda_{j0} \left[ \sum_{i = 1}^n p_i (2Z_i - 1) c_j(\mathbf{X}_i) \right]  \\
&\quad + \sum_{ j = 1 }^m \lambda_{j1} \left[ \sum_{i = 1}^n p_i Z_i c_j(\mathbf{X}_i) - \sum_{i = 1}^n q_ic_j(\mathbf{X}_i)  \right]
\end{split}
\end{equation}
where $\boldsymbol{\lambda}_0 \equiv (\lambda_{10}, \lambda_{20}, \ldots, \lambda_{m0})^{T}$ and $\boldsymbol{\lambda}_1 \equiv (\lambda_{11}, \lambda_{21}, \ldots, \lambda_{m1})^{T}$ with $\boldsymbol{\lambda} \equiv (\boldsymbol{\lambda}^{T}_0, \boldsymbol{\lambda}^{T}_1)^{T}$. The criterion distance function should be selected so that $P_f\left[q_i, \sum_{j = 1}^m c_j(\mathbf{X}_i)\lambda_{j0}\right] = \pi(\mathbf{X}_i)^{-1}$. 

We can also frame this problem using the notation from Section \ref{constrain}. Let $\mathbf{A}_0$ be an $(n\times m)$ matrix whose elements consist of $a_{ij0} = (2Z_i - 1)c_j(\mathbf{X}_i)$, $\mathbf{A}_1$ be an $(n\times m)$ with entries $a_{ij1} = Z_i c_j(\mathbf{X}_i)$, $\mathbf{b}_0 = \mathbf{0}_m$, and $\mathbf{b}_1$ be an $(m \times 1)$ vector with entries $b_{j1} = \sum_{i = 1}^n q_ic_j(\mathbf{X}_i)$. We then combine $\mathbf{A}_0$ and $\mathbf{A}_1$ to construct $\mathbf{A} = \left[\mathbf{A}_0, \mathbf{A}_1 \right]$ while $\mathbf{b}_0$ and $\mathbf{b}_1$ are concatenated into $\mathbf{b} = \left[\mathbf{b}^{T}_0, \mathbf{b}^{T}_1 \right]^{T}$.

After differentiating (\ref{lagrange-ate}) with respect to $p_i$ for some $i = 1,2,\ldots,n$ and setting the resulting derivative to zero, we arrive at the generalized projection evaluated at $\boldsymbol{\lambda} \in \Re^{2m}$. The dual objective function is obtained by substituting the generalized projection for $p_i$ in the Lagrangian. The dual solution solves the dual problem,
\begin{equation}\label{dual-ate}
\hat{\boldsymbol{\lambda}} = \argmax_{\boldsymbol{\lambda} \in \Re^{2m}} \sum_{i = 1}^n L_{\text{ATE}}\left\{P_f\left[q_i, \sum_{j = 1}^m (2Z_i - 1) c_j(\mathbf{X}_i)\lambda_{j0} + \sum_{j = 1}^m Z_ic_j(\mathbf{X}_i)\lambda_{j1} \right], q_i, \boldsymbol{\lambda} \right\}.
\end{equation} 
As a result of Propositions \ref{global-min} and \ref{dual-thm}, the balancing weights are uniquely determined by the generalized projection evaluated at $\hat{\boldsymbol{\lambda}} \in \Re^{2m}$,
\begin{equation}\label{weights-ate} 
\hat{p}(\mathbf{X}_i) = P_f\left[q_i, \sum_{j = 1}^m (2Z_i - 1) c_j(\mathbf{X}_i)\hat{\lambda}_{j0} + \sum_{j = 1}^m Z_ic_j(\mathbf{X}_i)\hat{\lambda}_{j1} \right], \enskip i = 1,2,\ldots,n. 
\end{equation}
It is recommended, and often necessary (see the proofs to Theorems \ref{dr-ate} and \ref{dr-hte}), that one of the balance functions be an intercept - we will assume throughout that $c_1(\mathbf{X}) = 1$ for all $\mathbf{X}$. This constraint implies $\sum_{\{i : Z_i = 1\}} \hat{p}_i = \sum_{\{i : Z_i = 0\}} \hat{p}_i$.

\subsection{Balancing Weights for the ATT}\label{ATT}

Next we consider the problem of finding balancing weights to estimate $\tau_{\text{ATT}}$. This requires solving the primal problem to
\begin{equation}\label{primal-att}
\begin{split} 
\text{minimize} &\quad \sum_{i = 1}^n \infdiv{p_i}{q_i}  \\ 
\text{subject to} &\quad \sum_{i = 1}^n p_i(1 - Z_i)c_j(\mathbf{X}_i) = \sum_{i = 1}^n q_i Z_i c_j(\mathbf{X}_i) \enskip \text{for all} \enskip j = 1,2,\ldots,m. 
\end{split}
\end{equation} 
The criterion Bregman distance should be chosen so that the generalized projection resembles the functional form for the odds of treatment. That is, \[ P_f\left[q_i, \sum_{j = 1}^m c_j(\mathbf{X}_i)\lambda_j\right] = \frac{\pi(\mathbf{X}_i)}{1 - \pi(\mathbf{X}_i)} \] where $\boldsymbol{\lambda} \equiv (\lambda_1, \lambda_2, \ldots, \lambda_m)^{T}$. In terms of the notation presented in Section \ref{constrain}, (\ref{primal-att}) is equivalent to (\ref{primal}) by setting $a_{ij} = (1 - Z_i)c_j(\mathbf{X}_i)$, $i = 1,2,\ldots,n$, and $b_j = \sum_{i = 1}^n q_i Z_i c_j(\mathbf{X}_i)$, $j = 1,2,\ldots,m$. Similar to the balancing weights for estimating $\tau_{\text{ATE}}$, we set $c_1(\mathbf{X}) = 1$ for all $\mathbf{X}$ so that $\sum_{\{i : Z_i = 1\}} q_i = \sum_{\{i : Z_i = 0\}} \hat{p}_i$. The associated Lagrangian for (\ref{primal-att}) can then be expressed as \[ L_{\text{ATT}}(\mathbf{p}, \mathbf{q}, \boldsymbol{\lambda}) \equiv \sum_{i = 1}^n \infdiv{p_i}{q_i} + \sum_{ j = 1 }^m \lambda_j \left[\sum_{i = 1}^n p_i (1 - Z_i) c_j(\mathbf{X}_i) - \sum_{i = 1}^n q_i Z_i c_j(\mathbf{X}_i) \right].\]

Notice that when $Z = 1$ and $q \in \Delta$, then $P_f\left[q, (1 - Z) \sum_{j = 1}^m c_j(\mathbf{X})\lambda_j \right] = q$. Moreover, recall from Section \ref{breg-distances} that $\infdiv{q}{q} = 0$ for some $q \in \Delta$. Therefore, (\ref{primal-att}) can be reconstructed into the equivalent primal problem to
\[ \begin{split} 
\text{minimize} &\quad \sum_{i = 1}^n \infdiv{p_i}{q_i}  \\
\text{subject to} &\quad \sum_{i = 1}^n p_i(2Z_i - 1)c_j(\mathbf{X}_i) = 0 \enskip \text{for all} \enskip j = 1,2,\ldots,m \enskip \text{and} \\
&\quad \ p_i = q_i \enskip \text{for all} \enskip i \in \{i : Z_i = 1\}.
\end{split} \]
According to Propositions \ref{global-min} and \ref{dual-thm}, the balancing weights are evaluated with the resulting generalized projection, \[ \hat{p}(\mathbf{X}_i) = P_f\left[q_i, (1-Z_i)\sum_{j = 1}^m c_j(\mathbf{X}_i)\hat{\lambda}_j \right], \enskip i = 1,2,\ldots,n, \] where the dual vector is estimated by solving for \[ \hat{\boldsymbol{\lambda}} = \argmax_{\boldsymbol{\lambda} \in \Re^m} \sum_{i = 1}^n L_{\text{ATT}}\left\{P_f\left[q_i, (1 - Z_i)\sum_{j = 1}^m c_j(\mathbf{X}_i) \lambda_j\right], q_i, \boldsymbol{\lambda}\right\}. \]

\section{Relationship with other Covariate Balance Methods}\label{related}

\subsection{Entropy Balancing}\label{ebal}

Entropy balancing \citep{hainmueller_entropy_2012} is a special case of a constrained Bregman distance optimization problem. By setting $f(\mathbf{p}) = \sum_{i = 1}^n p_i \log(p_i)$  and $q_i \in (0,\infty)$ for all $i = 1,2,\ldots,n$, we can identify the entropy balancing primal problem, which is to
\begin{equation}\label{primal-ebal}
\begin{split} 
\text{minimize} &\quad \sum_{i = 1}^n \left[p_i\log\left(\frac{p_i}{q_i}\right) - p_i + q_i\right] \\ 
\text{subject to} &\quad \sum_{i = 1}^n p_i(1 - Z_i)c_j(\mathbf{X}_i) = \sum_{i = 1}^n q_i Z_i c_j(\mathbf{X}_i) \enskip \text{for all} \enskip j = 1,2,\ldots,m. 
\end{split}
\end{equation} 
According to Proposition \ref{dual-thm}, optimizing (\ref{primal-ebal}) is equivalent to maximizing the dual objective function,
\begin{equation}\label{dual-ebal} 
\hat{\boldsymbol{\lambda}} = \argmax_{\boldsymbol{\lambda} \in \Re^m} \ \sum_{i = 1}^n \left\{ -q_i\exp\left[-(1-Z_i)\sum_{j = 1}^m c_j(\mathbf{X}_i) \lambda_j\right] - q_i Z_i \sum_{j = 1}^m c_j(\mathbf{X}_i) \lambda_j \right\} .
\end{equation} 
The vector of balancing weights is obtained by evaluating the generalized projection with the solution to the dual problem, which yields 
\begin{equation}\label{weights-ebal}  
\hat{p}(\mathbf{X}_i) = q_i\exp\left[-(1 - Z_i)\sum_{j = 1}^m c_j(\mathbf{X}_i) \hat{\lambda}_j\right], \enskip i = 1,2,\ldots,n. 
\end{equation}
In \cite{hainmueller_entropy_2012}, (\ref{primal-ebal}) is written using the normalized relative entropy instead of the unnormalized relative entropy. However, optimizing the normalized relative entropy is simply achieved by (\ref{dual-ebal}) and (\ref{weights-ebal}) with minor alterations. Let $q'_i = q_i/\sum_{i = 1}^n q_iZ_i$ for all $i \in \{i:Z_i = 1\}$ and constrain the intercept so that $\sum_{i = 1}^n \hat{p}_i(1 - Z_i) = \sum_{i = 1}^n q'_i Z_i = 1$.  As previously suggested in Section \ref{ATT}, we recommend setting $c_1(\mathbf{X}) = 1$. In doing so, the resulting balancing weights for the control group will sum to one while still satisfying the constraints of the primal problem.

Using the resulting estimating equations for $\boldsymbol{\lambda}$ and $\tau_{\text{ATT}}$ in concordance with results from M-estimation theory \citep{stefanski_calculus_2002}, \cite{zhao_entropy_2017} show that entropy balancing weights produce doubly-robust estimates of $\tau_{\text{ATT}}$. This means if either $\text{logit}[\pi(\mathbf{X})] \in \text{span}\{c_j(\mathbf{X}):j = 1,2,\ldots,m\}$ or $\mu_0(\mathbf{X}) \in \text{span}\{c_j(\mathbf{X}):j = 1,2,\ldots,m\}$, then the balancing weights of (\ref{weights-ebal}) applied to (\ref{HT}) is consistent for $\tau_{\text{ATT}}$. If both conditions are satisfied, then the estimator achieves the semiparametric efficiency bound derived by \cite{hahn_role_1998} for estimators of $\tau_{\text{ATT}}$. The Horvitz-Thompson estimator for $\tau_{\text{ATT}}$ that substitutes a consistent estimate of the propensity score for $\pi(\mathbf{X})$ in (\ref{HT-ATT}), on the other hand, does not achieve the semiparametric efficiency bound.

\subsection{Covariate Balance Propensity Scores}\label{cbps}

Another method for covariate balance, developed by \cite{imai_covariate_2013}, proposes fitting a logit model for the propensity score,
\begin{equation}\label{logit-model} 
\pi(\mathbf{X}_i) = \frac{\exp\left[\sum_{j = 1}^m c_j(\mathbf{X}_i)\lambda_j\right]}{1 + \exp\left[\sum_{j = 1}^m c_j(\mathbf{X}_i)\lambda_j\right]}, \enskip i = 1,2,\ldots,n,
\end{equation} 
subject to 
\begin{equation}\label{logit-constraint} 
\sum_{i = 1}^n \left[\frac{Z_{i}c_j(\mathbf{X}_i)}{\pi(\mathbf{X}_i)} - \frac{(1 - Z_{i})c_j(\mathbf{X}_i)}{1 - \pi(\mathbf{X}_i)}\right] = 0 \enskip \text{for all} \enskip j = 1,2,\ldots,m. 
\end{equation} 
They opted to solve for $\hat{\boldsymbol{\lambda}} \in \Re^m$ using generalized method of moments (GMM) while at the same time satisfying (\ref{logit-constraint}). The estimated propensity scores can be transformed into balancing weights for estimating $\tau_{\text{ATE}}$ with the inverse probability of treatment weighting estimator. We will refer to the model where the balance functions that appear in (\ref{logit-constraint}) are identical to the balance functions within the linear predictor of (\ref{logit-model}) as the exactly-specified CBPS model.

The weights obtained with an exactly-specified CBPS model can be expressed in an equivalent manner to a constrained optimization problem following our framework. First, notice that the fixed effect coefficients of the logit model can double as a vector of dual variables. Next, observe that (\ref{logit-constraint}) can be rewritten as \[ \sum_{i = 1}^n \left\{1 + \exp\left[-(2Z_i - 1) \sum_{j = 1}^m c_j(\mathbf{X}_i) \lambda_j \right]\right\} (2Z_i - 1)c_j(\mathbf{X}_i) = 0 \enskip \text{for all} \enskip j = 1,2,\ldots,m. \] The CBPS primal problem can then be constructed in order to
\begin{equation}\label{sent}
\begin{split} 
\text{minimize} &\quad \sum_{i = 1}^n \left[(p_i - 1)\log\left(\frac{p_i - 1}{q_i - 1}\right) - p_i + q_i \right] \\
\text{subject to} &\quad \sum_{i = 1}^n p_i(2Z_i - 1)c_j(\mathbf{X}_i) = 0 \enskip \text{for all} \enskip j = 1,2,\ldots,m.
\end{split}
\end{equation} 
We call the criterion distance function in (\ref{sent}) the shifted relative entropy which is generated by setting $f(\mathbf{p}) = \sum_{i = 1}^n (p_i - 1)\log(p_i - 1)$, $\mathbf{p} \in [1,\infty)^n$. Notice that (\ref{sent}) also contains fewer constraints than (\ref{primal-ate}). We assume $q_i = 2$ for all $i = 1,2,\ldots,m$. Assuming uniform sampling weights follows the prevailing philosophy of the causal inference literature in which observational data are typically randomly sampled from the population of interest. The solution to the dual problem for (\ref{sent}) finds
\begin{equation}\label{dual-cbps}
\hat{\boldsymbol{\lambda}} = \argmax_{\boldsymbol{\lambda} \in \Re^m} \ \sum_{i = 1}^{n} \left\{ (2Z_i - 1)\sum_{j = 1}^m  c_j(\mathbf{X}_i)\lambda_j - \exp\left[-(2Z_i - 1)\sum_{j = 1}^m c_j(\mathbf{X}_i)\lambda_j \right]\right\}.
\end{equation} 
The principal reason for selecting the shifted relative entropy as the criterion distance function are the resulting balancing weights which resemble the inverse probability of treatment weights,
\begin{equation}\label{weights-cbps}
\hat{p}(\mathbf{X}_i) = 1 + \exp\left[-(2Z_i - 1)\sum_{j = 1}^m  c_j(\mathbf{X}_i)\hat{\lambda}_j \right], \enskip i = 1,2,\ldots,n. 
\end{equation} 
A similar derivation of CBPS using the dual function setup was also described by \cite{zhao_covariate_2019}. 

\cite{fan_improving_2018} identifies a condition that the balance functions must satisfy in order for CBPS to be doubly-robust for estimating $\tau_\text{ATE}$. This condition is not obvious from a data analytic context. However, the condition is satisfied if we assume a constant conditional average treatment effect. Under this assumption, we can prove that CBPS is doubly-robust using the balancing weights produced by (\ref{weights-cbps}).

\begin{assumption}[Constant Conditional ATE]\label{constant} For all $\mathbf{X}$, $\mu_1(\mathbf{X}) - \mu_0(\mathbf{X}) = \tau$. \end{assumption}

\begin{theorem}\label{dr-ate}
Let Assumptions \ref{sita} and \ref{positivity} be given. Suppose $\mathbb{E}[Y(0)]$, $\mathbb{E}[Y(1)]$, and $\mathbb{E}[c_j(\mathbf{X})]$ exist for all $j = 1,2,\ldots, m$. Furthermore, assume $\mathbb{V}[Y(0)] < \infty$ and $\mathbb{V}[Y(1)]  < \infty$. Then the balancing weights determined by (\ref{dual-cbps}) and (\ref{weights-cbps}) applied to (\ref{HT}) is doubly-robust in the sense that:
\begin{enumerate}
\item If $\text{logit}[\pi(\mathbf{X})] = \sum_{j = 1}^m c_j(\mathbf{X})\lambda_j$ for some $\lambda_j \in \Re$, $j = 1,2,\ldots,m$, then $\hat{\tau}$ is consistent for $\tau_{\text{ATE}}$;
\item Under Assumption \ref{constant} and if $\mu_0(\mathbf{X}) = \sum_{j = 1}^m c_j(\mathbf{X})\beta_j$ for some $\beta_j \in \Re$, $j = 1,2,\ldots,m$, then $\hat{\tau}$ is consistent for $\tau_{\text{ATE}}$;
\item If conditions 1 and 2 are both satisfied, then \[ \sqrt{n}(\hat{\tau} - \tau_{\text{ATE}}) \rightarrow_{d} \mathcal{N}(0, \Sigma_{\text{semi}}) \]
where \[ \Sigma_{\text{semi}} = \mathbb{E}\left\{\frac{\mathbb{V}[Y(1)|\mathbf{X}]}{\pi(\mathbf{X})} + \frac{\mathbb{V}[Y(0)|\mathbf{X}]}{1 - \pi(\mathbf{X})}\right\}. \]
\end{enumerate} 
\end{theorem}

As an extension to CBPS, consider the primal problem using the Bregman distance generated by setting $f(\mathbf{p}) = \sum_{i = 1}^n p_i\log(p_i) + (1-p_i)\log(1-p_i)$,  $\mathbf{p} \in [0,1]^n$:
\begin{equation}\label{bent}
\begin{split} 
\text{minimize} &\quad \sum_{i = 1}^n \left[p_i\log\left(\frac{p_i}{q_i}\right) + (1 - p_i)\log\left(\frac{1 - p_i}{1 - q_i}\right)\right] \\
\text{subject to} &\quad \sum_{i = 1}^n p_i(2z_i - 1)c_j(\mathbf{X}_i) = 0 \enskip \text{for all} \enskip j = 1,2,\ldots,m.
\end{split}
\end{equation} 
If we assume $q_i = 1/2$ for all $i = 1,2,\ldots,n$, then according to Propositions \ref{global-min} and \ref{dual-thm} the solution to (\ref{bent}) is
\begin{equation}\label{weights-owate}
\hat{p}(\mathbf{X}_i) = \frac{1}{1 + \exp\left[(2Z_i - 1)\sum_{j = 1}^m c_j(\mathbf{X}_i)\hat{\lambda}_j \right]},
\end{equation} 
where the dual solution is obtained by solving for
\begin{equation}\label{dual-owate}
\begin{split}
\hat{\boldsymbol{\lambda}} = \argmax_{\boldsymbol{\lambda} \in \Re^m } & \sum_{i = 1}^n \frac{1}{1 + \exp\left[(2Z_i - 1) \sum_{j = 1}^m c_j(\mathbf{X}_i)\lambda_j\right]} \log\left[ \frac{2}{1 + \exp\left[(2Z_i - 1)\sum_{j = 1}^m c_j(\mathbf{X}_i)\lambda_j\right]} \right] \\
&+  \sum_{i = 1}^n \frac{1}{1 + \exp\left[-(2Z_i - 1) \sum_{j = 1}^m c_j(\mathbf{X}_i)\lambda_j\right]}\log\left[ \frac{2}{1 + \exp\left[-(2Z_i - 1)\sum_{j = 1}^m c_j(\mathbf{X}_i)\lambda_j\right]} \right] \\
&+ \sum_{j = 1}^m \lambda_j \left\{ \sum_{i = 1}^n \frac{(2Z_i - 1)c_j(\mathbf{X}_i)}{1 + \exp\left[(2Z_i - 1)\sum_{j = 1}^m c_j(\mathbf{X}_i)\lambda_j\right]} \right\}.
\end{split}
\end{equation} 
The Bregman distance in this case is referred to as the binary relative entropy. This distance is useful for finding balancing weights that produce estimates for a special case of the weighted average treatment effect called the optimally weighted average treatment effect (OWATE) \citep{crump_moving_2006}, 
\[ \tau_{\text{OWATE}} \equiv \frac{\mathbb{E}\{\pi(\mathbf{X})[1 - \pi(\mathbf{X})][Y(1) - Y(0)]\}}{\mathbb{E}\{\pi(\mathbf{X})[1 - \pi(\mathbf{X})]\}}. \] A consistent estimator for $\tau_{\text{OWATE}}$ is also consistent for $\tau_{\text{ATE}}$, with the smallest variance, when we are given Assumption \ref{constant} and the potential outcomes have equal variance. \cite{li_balancing_2018} further motivates the use of estimators for $\tau_{\text{OWATE}}$ when there is poor overlap between the treated and control groups. Equations (\ref{weights-owate}) and (\ref{dual-owate}) provide a dual interpretation of the covariate balance scoring rule for estimating $\tau_{\text{OWATE}}$ considered by \cite{zhao_covariate_2019}. By replacing $p(\mathbf{X})$ with (\ref{weights-owate}) in (\ref{HT}), and using arguments similar to the proof of Theorem 1, we derive a doubly-robust estimator for $\tau_{\text{OWATE}}$ with the usual asymptotic properties.

\begin{corollary}\label{dr-owate}
Under the same assumptions and conditions as Theorem \ref{dr-ate}, the balancing weights determined by (\ref{weights-owate}) and (\ref{dual-owate}) applied to (\ref{HT}) is doubly-robust for estimating $\tau_{\text{OWATE}}$ with asymptotic variance
\begin{equation}
\Sigma_{\text{semi}} = \frac{\mathbb{E}\left(\pi(\mathbf{X})^2[1 - \pi(\mathbf{X})]^2 \left\{\frac{\mathbb{V}[Y(1)|\mathbf{X}]}{\pi(\mathbf{X})} + \frac{\mathbb{V}[Y(0)|\mathbf{X}]}{1 - \pi(\mathbf{X})}\right\}\right)}{\mathbb{E}\{\pi(\mathbf{X}) [1 - \pi(\mathbf{X})]\}^2}.
\end{equation}
\end{corollary}

\subsection{Improved Covariate Balance Propensity Scores}

The iCBPS approach \citep{fan_improving_2018} improves upon the CBPS method described in Section \ref{cbps} to better accommodate heterogeneous treatment effects. The objective of this method is to fit a logit model of the propensity score subject to the constraints
\begin{equation}\label{logit-constraint-2} 
\begin{split}
\sum_{i = 1}^n \left[\frac{Z_{i}c_j(\mathbf{X}_i)}{\pi(\mathbf{X}_i)} - \frac{(1 - Z_{i})c_j(\mathbf{X}_i)}{1 - \pi(\mathbf{X}_i)}\right] &= 0 \enskip \text{and} \\
\sum_{i = 1}^n \left[\frac{Z_{i}}{\pi(\mathbf{X}_i)} - 1\right] c_j(\mathbf{X}_i) &= 0 \enskip \text{for all} \enskip j = 1,2,\ldots,m.
\end{split}
\end{equation} 
\cite{fan_improving_2018} uses GMM to estimate $\hat{\boldsymbol{\lambda}} \in \Re^m$ in (\ref{logit-model}) subject to (\ref{logit-constraint-2}). This modified approach can be adapted to fit into our proposed framework, with the balancing weights being estimated using dual optimization techniques instead of GMM.

Using the same criterion Bregman distance as the one used in (\ref{sent}), we can obtain balancing weights that satisfy (\ref{logit-constraint-2}) as follows. Assume $q_i = 2$ for all $i = 1,2,\ldots,m$. Define the primal problem for iCBPS as
\begin{equation}\label{primal-icbps}
\begin{split} 
\text{minimize} &\quad \sum_{i = 1}^{n} \left[(p_i - 1) \log\left(p_i - 1\right) - p_i + 2\right] \\
\text{subject to} &\quad \sum_{i = 1}^n p_i(2Z_i - 1)c_j(\mathbf{X}_i) = 0 \enskip \text{and} \\
&\quad \sum_{i = 1}^n p_i Z_i c_j(\mathbf{X}_i) = \sum_{i = 1}^n c_j(\mathbf{X}_i) \enskip \text{for all} \enskip j = 1,2,\ldots,m.
\end{split}
\end{equation}
As opposed to (\ref{sent}), the iCBPS primal problem follows our guidelines in Section \ref{ATE} more closely. The resulting dual solution solves for
\begin{equation}\label{dual-icbps}
\begin{split}
\hat{\boldsymbol{\lambda}} &= \argmax_{\boldsymbol{\lambda} \in \Re^{2m}} \ \sum_{i = 1}^{n} \Bigg\{ (2Z_i - 1)\sum_{j = 1}^m  c_j(\mathbf{X}_i)\lambda_{j0} + Z_i \sum_{j = 1}^m c_j(\mathbf{X}_i)\lambda_{j1} - \sum_{j = 1}^m c_j(\mathbf{X}_i)\lambda_{j1} \\
&\quad\quad - \exp\left[-(2Z_i - 1)\sum_{j = 1}^m c_j(\mathbf{X}_i)\lambda_{j0} -  Z_i \sum_{j = 1}^m  c_j(\mathbf{X}_i)\lambda_{j1} \right]\Bigg\}.
\end{split}
\end{equation} 
The covariate balancing weights differ slightly from (\ref{weights-cbps}) due to the additional constraints in (\ref{primal-icbps}) with
\begin{equation}\label{weights-icbps}
\hat{p}(\mathbf{X}_i) = 1 + \exp\left[-(2Z_i - 1)\sum_{j = 1}^m  c_j(\mathbf{X}_i)\hat{\lambda}_{j0} -  Z_i \sum_{j = 1}^m  c_j(\mathbf{X}_i)\hat{\lambda}_{j1} \right], \enskip i = 1,2,\ldots,n. 
\end{equation} 
Note that with the GMM approach, $\boldsymbol{\lambda} \in \Re^m$ whereas with our method, $\boldsymbol{\lambda} \in \Re^{2m}$. This implies that exact balance between covariates is not necessarily achieved with the proposed methods of \cite{fan_improving_2018}. A notable deviation from our own recommendations exists within how $\mathbf{b}$ is specified. Even though $q_i = 2$ for all $i = 1,2,\ldots,n$, we set $b_{j1} = \sum_{i = 1}^n c_j(\mathbf{X}_i)$ for all $j = 1,2,\ldots,m$. If we were to follow the setup in Section \ref{ATE}, we would set $b_{j1} = \sum_{i = 1}^n 2c_j(\mathbf{X}_i)$. However, this distinction is minor in the context of uniform sampling weights and should produce similar results. 

We now show that the weights produced by (\ref{dual-icbps}) and (\ref{weights-icbps}) applied to (\ref{HT}) is doubly-robust given a linear conditional average treatment effect, defined in Assumption \ref{linear}. Note that Assumption \ref{linear} is less stringent than Assumption \ref{constant} which was necessary to prove Theorem \ref{dr-ate}.

\begin{assumption}[Linear Conditional ATE]\label{linear} For all $\mathbf{X}$, $\mu_1(\mathbf{X}) - \mu_0(\mathbf{X}) = \sum_{j = 1}^m c_j(\mathbf{X}) \alpha_j$ where $\alpha_j \in \Re$ for all $j = 1,2,\ldots,m$. \end{assumption}

\begin{theorem}\label{dr-hte}
Let Assumptions \ref{sita} and \ref{positivity} be given. Suppose $\mathbb{E}[Y_i(0)]$, $\mathbb{E}[Y(1)]$ and $\mathbb{E}[c_j(\mathbf{X})]$ exist for all $j = 1,2,\ldots, m$. Assume $\mathbb{V}[Y(0)] < \infty$ and $\mathbb{V}[Y(1)] < \infty$. Then the balancing weights determined by (\ref{dual-icbps}) and (\ref{weights-icbps}) applied to (\ref{HT}) is doubly-robust in the sense that:
\begin{enumerate}
\item If $\text{logit}[\pi(\mathbf{X})] = \sum_{j = 1}^m c_j(\mathbf{X})\lambda_{j0}$ for some $\lambda_{j0} \in \Re$, $j = 1,2,\ldots,m$, then $\hat{\tau}$ is consistent for $\tau_{\text{ATE}}$;
\item Under Assumption \ref{linear} and if $\mu_0(\mathbf{X}) = \sum_{j = 1}^m c_j(\mathbf{X})\beta_j$ for some $\beta_j \in \Re$, $j = 1,2,\ldots,m$, then $\hat{\tau}$ is consistent for $\tau_{\text{ATE}}$;
\item If conditions 1 and 2 are both satisfied, then \[ \sqrt{n}(\hat{\tau} - \tau_{\text{ATE}}) \rightarrow_{d} \mathcal{N}(0, \Sigma_{\text{semi}}) \]
where \[ \Sigma_{\text{semi}} = \mathbb{E}\left\{\frac{\mathbb{V}[Y(1)|\mathbf{X}]}{\pi(\mathbf{X})} + \frac{\mathbb{V}[Y(0)|\mathbf{X}]}{1 - \pi(\mathbf{X})} + \left[\mu_1(\mathbf{X}) - \mu_0(\mathbf{X}) - \tau_{\text{ATE}}\right]^2\right\}. \]
\end{enumerate} 
\end{theorem}

\subsection{Calibration Estimators}\label{calib}

\cite{chan_globally_2015} describes a class of estimators originally introduced by \cite{deville_calibration_1992} for survey sampling called calibration estimators. One of the contributions from \cite{chan_globally_2015} shows how calibration estimators can be applied to covariate balance problems. For some separable generalized distance function $G:\Delta^n \rightarrow \Re$, calibration estimators find balancing weights that solve the primal problem to
\begin{equation}\label{primal-calib}
\begin{split} 
\text{minimize} &\quad \sum_{i = 1}^n G(p_i) \\ 
\text{subject to} &\quad \sum_{i = 1}^n p_i(1 - Z_i)c_j(\mathbf{X}_{i}) = \sum_{i = 1}^n c_j(\mathbf{X}_{i}) \enskip \text{and} \\ 
&\quad \sum_{i = 1}^n p_i Z_i c_j(\mathbf{X}_{i}) = \sum_{i = 1}^n c_j(\mathbf{X}_{i}) \enskip \text{for all} \ j = 1,2,\ldots,m.
\end{split}
\end{equation}
\cite{chan_globally_2015} assume uniform sampling weights. Equation (\ref{primal-calib}) is solved by defining the functions $h(p) \equiv G(1 - p)$ and $g(v) \equiv h\left[(\nabla h)^{-1}(v)\right] + v  - v (\nabla h)^{-1}(v)$ with $p \in \Delta$ and $v \in \Re$ to write the dual objective functions, which we use to solve for
\begin{equation}\label{dual-calib} 
\begin{split}
\hat{\boldsymbol{\lambda}}_0 &= \argmax_{\boldsymbol{\lambda} \in \Re^m} \ \sum_{i = 1}^n \left\{ g\left[\sum_{j = 1}^m (1 - Z_i) c_j(\mathbf{X}_i) \lambda_j\right] - \sum_{j = 1}^{m} c_j(\mathbf{X}_i)\lambda_j\right\} \enskip \text{and} \\ 
\hat{\boldsymbol{\lambda}}_1 &= \argmax_{\boldsymbol{\lambda} \in \Re^m} \ \sum_{i = 1}^n \left\{ g\left[\sum_{j = 1}^m  Z_i c_j(\mathbf{X}_i)\lambda_j\right] - \sum_{j = 1}^{m} c_j(\mathbf{X}_i)\lambda_j\right\}.
\end{split}
\end{equation}
The resulting balancing weights are obtained by evaluating the first derivative of $g$ at $\hat{\boldsymbol{\lambda}}_0$ and $\hat{\boldsymbol{\lambda}}_1$,
\begin{equation}\label{weights-calib} 
\hat{p}(\mathbf{X}_i) = \nabla g\left[\sum_{j = 1}^m (1 - Z_i) c_j(\mathbf{X}_i)\hat{\lambda}_{j0} + \sum_{j = 1}^m Z_i c_j(\mathbf{X}_i)\hat{\lambda}_{j1}\right]. 
\end{equation}

Similar to our solution for finding balancing weights to estimate $\tau_{\text{ATE}}$, the dual variable $\boldsymbol{\lambda} \equiv (\boldsymbol{\lambda}_0, \boldsymbol{\lambda}_1)^{T}$ has $2m$ entries. \cite{tseng_relaxation_1987} and \cite{chan_globally_2015} show that (\ref{dual-calib}) can be solved using any strictly concave $g(v)$, $v \in \Re$, assuming that a feasible solution for (\ref{primal-calib}) exists. Therefore, calibration estimators are not necessarily restricted to Bregman distances. However, if $G$ is a monotone increasing transformation of $D_f$ with respect to $\mathbf{p} \in \Delta^n$, then (\ref{primal-calib}) can be constructed using Bregman distances so that the primal solutions are equivalent. The only difference from the methods we present in Section \ref{ATE} is with the construction of $\mathbf{A}$ and $\mathbf{b}$. In Theorem \ref{equiv-hte} we identify the conditions for which our proposed method is equivalent to the calibration estimator approach of \cite{chan_globally_2015}. 

\begin{theorem}\label{equiv-hte}
Suppose we have a generalized distance $G(\mathbf{p})$ which is a monotone increasing transformation of some Bregman distance $\infdiv{\mathbf{p}}{\mathbf{q}}$ with respect to $\mathbf{p} \in \Delta^n$ and $\mathbf{q} \in \Delta^n$ is uniform. Then for 
\[ \begin{split} 
\Omega_0 &= \left\{\mathbf{p} : \sum_{i = 1}^n p_iZ_ic_j(\mathbf{X}_i) = b_j \ \text{and} \  \sum_{i = 1}^n p_i(1 - Z_i)c_j(\mathbf{X}_i) = b_j \ \text{for all} \ j = 1,2\ldots, m\right\} \enskip \text{and} \\ 
\Omega_1 &= \left\{ \mathbf{p} : \sum_{i = 1}^n p_iZ_ic_j(\mathbf{X}_i) = b_j \ \text{and} \  \sum_{i = 1}^n p_i Z_ic_j(\mathbf{X}_i) =  \sum_{i = 1}^n p_i (1 - Z_i)c_j(\mathbf{X}_i) \ \text{for all} \ j = 1,2\ldots,m \right\},
\end{split} \] 
$\tilde{\mathbf{p}} = \hat{\mathbf{p}}$ where $\tilde{\mathbf{p}} = \argmin_{\mathbf{p} \in \Omega_0 \cap \Delta^n} G(\mathbf{p})$ and $\hat{\mathbf{p}} = \argmin_{\mathbf{p} \in \Omega_1 \cap \Delta^n} \infdiv{\mathbf{p}}{\mathbf{q}}$.
\end{theorem}

Theorem 1 of \cite{chan_globally_2015} shows that calibration estimators can be used to produce consistent estimates for $\tau_{\text{ATE}}$ while also attaining the semiparametric efficiency bound described by \cite{hahn_role_1998}. This is accomplished using a nonparametric setup where the balance functions represent a basis for uniformly approximating $\mu_0(\mathbf{X})$, $\mu_1(\mathbf{X})$, and $\pi(\mathbf{X})$. Given this result and Theorem \ref{equiv-hte} implies that for a sufficiently rich set of balance functions, the Bregman distance weights in conjunction with (\ref{HT}) can produce consistent and efficient estimates of $\tau_{\text{ATE}}$. This result is quite useful when the balance functions that determine either the outcome or the treatment assignment are unknown.

Without further defining the distance to be optimized in the primal problem, (\ref{dual-calib}) and (\ref{weights-calib}) are less flexible when considering non-uniform sampling weights. This is especially important when developing iterative estimation algorithms or dealing with more complex balance designs where the data are not sampled uniformly from the population of interest. Furthermore, calibration estimators, as they are described in \cite{chan_globally_2015}, achieve a three-way balance between the treated, the controls, and the combined treatment groups for estimating $\tau_{\text{ATE}}$. As shown in Theorem \ref{dr-ate}, this condition is not required when Assumption \ref{constant} holds. \cite{zhao_covariate_2019} also noted that this condition is not required to achieve global efficiency using covariate balance scoring rules.

\section{Numerical Studies}\label{numerical}

\subsection{Homogeneous Treatment Effect Simulation}\label{homogeneous}

In this section, we demonstrate the utility of the proposed methodology using simulated data that assumes a constant conditional average treatment effect (Assumption \ref{constant}). We generate $1000$ replications of several datasets determined by one of $72$ experimental scenarios. For each dataset, we find balancing weights from four different covariate balancing methods to estimate $\tau_{\text{ATE}}$. They are:
\begin{enumerate}
\item (IPW) Inverse probability of treatment weights where the propensity score follows a logit model fit using maximum likelihood estimation;
\item (CBPS) Inverse probability of treatment weights where the propensity score is fit to an exactly-specified logit model subject to (\ref{logit-constraint}). The propensity scores are fit using generalized method of moments as implemented in the \texttt{CBPS} package \citep{fong_cbps:_2018};
\item (SENT) Balancing weights that are estimated by minimizing the shifted relative entropy following the results of (\ref{dual-icbps}) and (\ref{weights-icbps}). Using these balancing weights instead of (\ref{dual-cbps}) and (\ref{weights-cbps}) allows us to test the effect of over-specifying the linear constraints when we know Assumption \ref{constant} is satisfied;
\item (BENT) Balancing weights that are estimated by minimizing the binary relative entropy subject to the constraints in (\ref{bent}) via the dual and primal solutions of (\ref{weights-owate}) and (\ref{dual-owate}).
\end{enumerate}

We consider an extensive set of experimental scenarios adapted from those examined by \cite{kang_demystifying_2007}. These scenarios vary the sample size $n \in \{200,1000\}$, the error variance $\sigma^2 \in \{2,5,10\}$, the generative process that determines the treatment assignment (indexed by $\{a,b\}$), the outcome process (indexed by $\{a, b\}$), and the correlation between the potential outcomes, $\rho \in \{-0.3, 0, 0.5\}$. The covariates to be balanced (i.e the balance functions) are distributed as $X_{1},X_{2},X_{3},X_{4} \sim \mathcal{N}(0, 1)$. Define the transformations $U_1 = \exp(X_1/2)$, $U_2 = X_2/[1+\exp(X_1)] + 10$, $U_3 = (X_1X_3/25 + 0.6)^3$ and $U_4 = (X_2 + X_4 + 20)^2$. The vector $(U_1, U_2, U_3, U_4)^{T}$ is subsequently standardized to have a mean of zero and marginal variances of one.

The probability that a subject receives the treatment is then determined using the inverse logit link function, \[ \pi^{(k)}_i = \frac{\exp\left[\eta^{(k)}_i\right]}{1+\exp\left[\eta^{(k)}_i\right]}, \enskip k \in \{a, b\}.\]
Scenarios $a$ and $b$ distinguish whether the log odds of the propensity score is either linear or non-linear with
\begin{equation}\label{eta}
\begin{split}
\eta^{(a)}_{i} &= -X_{i1} + 0.5X_{i2} - 0.25X_{i3} - 0.1X_{i4} \enskip \text{and} \\
\eta^{(b)}_{i} &= -U_{i1} + 0.5U_{i2} - 0.25U_{i3} - 0.1U_{i4}. 
\end{split}
\end{equation}
The treatment indicators are generated by sampling $Z_{i} \sim \text{Bin}\left(1, \pi^{(k)}_{i}\right)$. For the outcome process, we use the bivariate model \[ \begin{bmatrix} Y_{i}(0) \\ Y_{i}(1) \end{bmatrix} \sim \mathcal{N}\left(\begin{bmatrix} \mu^{(\ell)}_i \\ \mu^{(\ell)}_i + \tau \end{bmatrix}, \begin{bmatrix} \sigma^2 & \rho \sigma^2 \\ \rho \sigma^2 & \sigma^2 \end{bmatrix}\right), \] where $\ell \in \{a, b\}$ indexes
\begin{equation}\label{mu}
\begin{split}
\mu^{(a)}_{i} &= 210 + 27.4X_{i1} + 13.7X_{i2} + 13.7X_{i3} + 13.7X_{i4} \enskip \text{and} \\
\mu^{(b)}_{i} &= 210 + 27.4U_{i1} + 13.7U_{i2} + 13.7U_{i3} + 13.7U_{i4}.
\end{split}
\end{equation}
Once the potential outcomes have been generated, the observed outcome is the potential outcome corresponding to the observed treatment assignment. Each of the covariate balancing methods listed above are provided the design matrix with an intercept and the four original covariates; $X_{i1}$, $X_{i2}$, $X_{i3}$, and $X_{i4}$ for $i = 1,2,\ldots,n$. The causal effect is then estimated using (\ref{HT}) where we substitute $p(\mathbf{X}_i)$ with the balancing weights estimated by each method.

\begin{figure}[H]
	\centering
	\includegraphics[scale = 0.6]{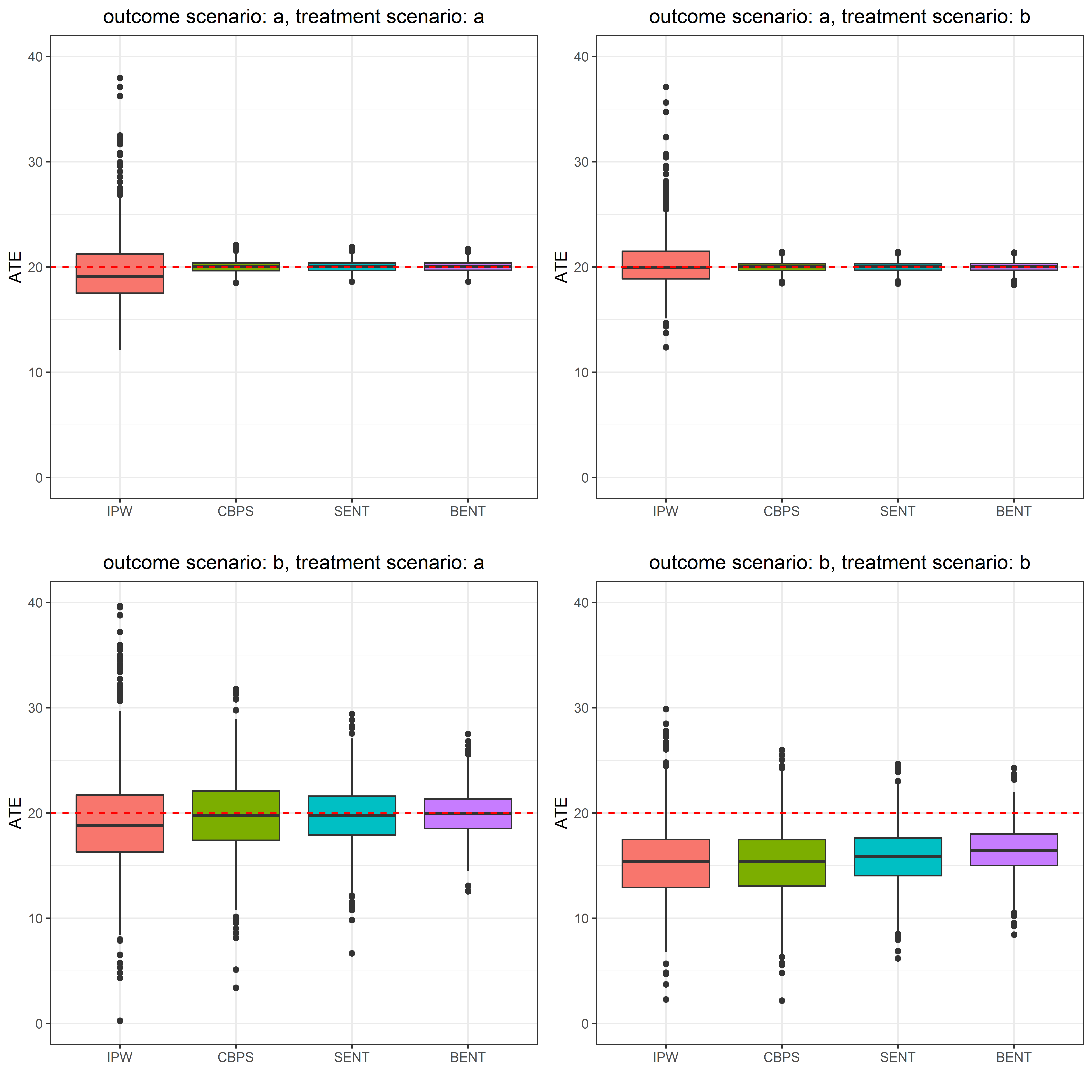}
	\caption{A subset of the constant conditional ATE estimates using four different methods for estimating balancing weights. Each boxplot is composed of $1000$ estimates from the replicates that generate the values in Table \ref{sim-table-1}.}\label{ATE-plot}
\end{figure}

We found that the correlation between the potential outcomes did not affect the resulting estimates of $\tau_{\text{ATE}}$. In addition, the effects of altering $\sigma^2$ and $n$ had anticipated results. Lower values of $\sigma^2$ led to lower standard errors of the causal effect estimate whereas smaller values of $n$ led to larger standard errors. Therefore, we report the results for $\rho = 0$, $n = 200$, and $\sigma^2 = 10$ in Table \ref{sim-table-1} and Figure \ref{ATE-plot}. The complete results appear in the online supplement.

For all the methods that we tested, if either the outcome model or the treatment assignment is correctly specified, then the causal effect estimate is unbiased. We see in Table \ref{sim-table-1} and Figure \ref{ATE-plot} that the balancing weights obtained with SENT perform as well, or better in some cases, than the exactly-specified CBPS model, even though the balancing weights obtained with SENT have twice as many constraints. The Monte Carlo standard error and bias of the estimates for $\tau_{\text{ATE}}$ are smallest when BENT is used to estimate the balancing weights for every scenario we examined. This is expected since these weights are used for estimating the $\tau_{\text{OWATE}}$ \citep{crump_moving_2006}, and because every condition necessary to ensure that an estimator for $\tau_{\text{OWATE}}$ is also an estimator for $\tau_{\text{ATE}}$ are met. The Monte Carlo standard errors and mean square error are also uniformly smaller for the average treatment effect estimates when using balancing weights estimated by CBPS and SENT versus IPW. This result indicates that methods which exactly balance the empirical covariate distributions perform better in finite sample settings. When both the outcome and treatment assignment models are misspecified, the four methods for finding balancing weights all produced biased estimates of $\tau_{\text{ATE}}$. In these completely misspecified scenarios, the balancing weights estimated with BENT produce the least amount of bias and the lowest standard error for estimating $\tau_{\text{ATE}}$.

\subsection{Heterogeneous Treatment Effect Simulation}\label{heterogeneous}

In this section we simulate an additional $72$ scenarios with a linear conditional average treatment effect to test our proposed methods under Assumption \ref{linear}. We use the same covariate distributions for $(X_{1},X_{2},X_{3},X_{4})^{T}$ and $(U_1, U_2, U_3, U_4)^{T}$ as in Section \ref{homogeneous}. We also recycle the conditional mean functions $\mu_i^{(\ell)}$ , $\ell \in \{a, b\}$, from (\ref{mu}). To generate the linear conditional average treatment effects, define 
\[ \begin{split}
\delta^{(a)}_{i} &= 20 - 13.7X_{i1} + 13.7X_{i4} \enskip \text{and} \\
\delta^{(b)}_{i} &= 20 - 13.7U_{i1} + 13.7U_{i4}.
\end{split} \]
For outcome scenarios $a$ and $b$, the bivariate outcome model is defined as \[ \begin{bmatrix} Y_{i}(0) \\ Y_{i}(1) \end{bmatrix} \sim \mathcal{N}\left(\begin{bmatrix} \mu^{(\ell)}_i \\ \mu^{(\ell)}_i + \delta^{(\ell)}_i \end{bmatrix}, \begin{bmatrix} \sigma^2 & \rho \sigma^2 \\ \rho \sigma^2 & \sigma^2 \end{bmatrix}\right), \enskip \ell \in \{a,b\}, \] from which we sample $n \in \{200, 1000\}$ entries. Each unit's treatment assignment is sampled from $\text{Bin}\left(1, \pi^{(k)}_{i}\right)$, where $\pi^{(k)}_{i}$ is determined by $\eta^{(k)}_{i}$, $k \in \{a,b\}$, which are defined in (\ref{eta}). Similar to the simulations conducted in Section \ref{homogeneous}, we also vary  $\sigma^2 \in \{2,5,10\}$ and $\rho \in \{-0.3, 0, 0.5\}$. For this set of scenarios we examine five different covariate balance methods:
\begin{enumerate}
\item (AIPW) Augmented inverse probability weights which uses the estimator 
\begin{equation}\label{AIPW} 
\hat{\tau}_{\text{AIPW}} = \frac{1}{n} \sum_{i = 1}^n \left\{ \frac{Z_iY_i}{\hat{\pi}(\mathbf{X}_i)} - \frac{[Z_i - \hat{\pi}(\mathbf{X}_i)]\hat{\mu}_1(\mathbf{X}_i)}{\hat{\pi}(\mathbf{X}_i)}  - \frac{(1 - Z_i)Y_i}{1 - \hat{\pi}(\mathbf{X}_i)} - \frac{[Z_i - \hat{\pi}(\mathbf{X}_i)]\hat{\mu}_0(\mathbf{X}_i)}{1 - \hat{\pi}(\mathbf{X}_i)}\right\}.
\end{equation} $\hat{\mu}_1(\mathbf{X})$ is fit using linear regression on the treated group and $\hat{\mu}_0(\mathbf{X})$ is fit using linear regression on the controls. $\hat{\pi}(\mathbf{X})$ is fit with logistic regression;
\item (CAL) Calibration weights that solve (\ref{dual-calib}) and (\ref{weights-calib}) for $g(v) = \exp(-v)$, $v \in \Re$. This is equivalent to minimizing the unnormalized relative entropy subject to the linear constraints in (\ref{primal-calib}). The R package \texttt{ATE} developed by \cite{haris_ate:_2015} is used to estimate these balancing weights;
\item (iCBPS) Inverse probability of treatment weights where the propensity scores follow an exactly-specified logit model which must satisfy (\ref{logit-constraint-2}) following the results of \cite{fan_improving_2018}. The propensity score is estimated using generalized method of moments as implemented in the \texttt{CBPS} package \citep{fong_cbps:_2018};
\item (hdCBPS) An augmented version of CBPS that extends (\ref{AIPW}) by using regularized regression techniques to debiased estimates of $\mu_1(\mathbf{X})$, and $\mu_0(\mathbf{X})$. The R package \texttt{CBPS} \citep{fong_cbps:_2018} is used to implement this method;
\item (SENT) Balancing weights that minimize the shifted relative entropy conditioned on the constraints in (\ref{primal-icbps}) where the estimates are obtained from (\ref{dual-icbps}) and (\ref{weights-icbps}).
\end{enumerate}
For CAL, iCBPS, and SENT, the resulting balancing weights are substituted for $p(\mathbf{X})$ in (\ref{HT}) to estimate $\tau_{\text{ATE}}$. The augmented approach of AIPW was first proposed by \cite{robins_estimation_1994} while hdCBPS uses the augmented estimator proposed by \cite{ning_robust_2018}.

\begin{figure}[H]
	\centering
	\includegraphics[scale = 0.6]{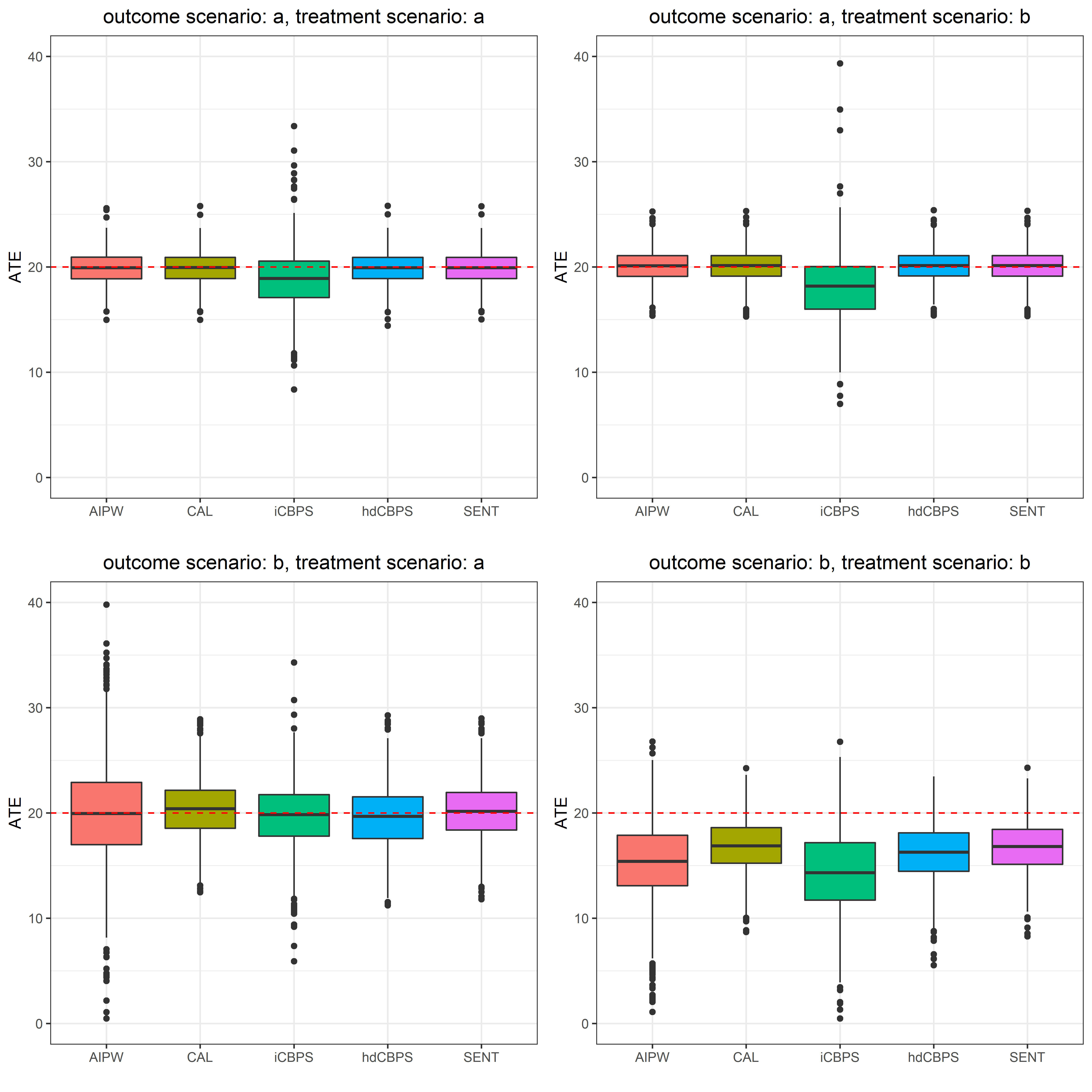}
	\caption{A subset of the linear conditional ATE estimates using five different methods for estimating balancing weights. Each boxplot is composed of $1000$ estimates from the replicates that generate the values in Table \ref{sim-table-2}.}\label{HTE-plot}
\end{figure}

As with the previous simulation study, it appears that the correlation between potential outcomes is inconsequential while the Monte Carlo standard errors predictably decrease when either $n$ increases or $\sigma^2$ decreases. A representative selection of results from the experiment where $\rho = 0$, $n = 200$, and $\sigma^2 = 10$ are found in Table \ref{sim-table-2} and Figure \ref{HTE-plot}. The complete results can be found in the online supplement. This simulation demonstrates that all five methods enjoy the doubly-robust property described in Theorem \ref{dr-hte}. For each scenario, CAL and SENT had similar levels of bias and variation despite using different criterion distance functions. Even though the constraints of iCBPS and SENT are the same, the differences between the optimization techniques of the two methods becomes quite apparent. The Monte Carlo standard error of the estimates for $\tau_{\text{ATE}}$ using balancing weights obtained with SENT is smaller than the standard error of the estimates using balancing weights found with iCBPS. AIPW, CAL, and SENT performed similarly whenever the outcome was correctly specified. However, when the outcome model is misspecified and the propensity score is correctly specified, the Monte Carlo standard error and mean squared error were greater with AIPW compared to SENT and CAL. This suggests, and is further confirmed by hdCBPS, that methods which exactly-balance covariate distributions can improve the efficiency of a doubly-robust estimator in finite samples. hdCBPS performed about as well as SENT and CAL. However, this method was proposed to alleviate issues encountered with high-dimensional covariate data, rendering many of its benefits redundant in this low-dimensional simulation study.

\subsection{Illustrative Example of Unplanned Readmissions after Lung Resection}\label{example}

Next, we investigate the results of a real data set using different weighting and matching methods. In \cite{bhagat_national_2017}, the odds of unplanned, 30-day readmissions are compared between lung cancer patients that receive thoracoscopic versus open lung resections. The study identified 9,510 patients that underwent some form of lung resection from the American College of Surgeons - National Safety and Quality Innovation Program (ACS-NSQIP) database. Of those 9,510 patients, 4,935 (51.9\%) received a thoracoscopic resection and 4,575 (48.1\%) received an open anatomic resection. The study analysis carried out a greedy one-to-one matching of patients using the estimated propensity score as the criterion matching function \citep{ho_matching_2007}. The propensity scores were fit with standard logistic regression. This algorithm matched 3,399 thoracoscopic lung resection patients to 3,399 open anatomic lung resection patients, dropping 2,712 patients (28.5\%). In doing so, the ``treated'' group are assumed to be the patients that receive thoracoscopic lung resections and represent a random sample of  the target population. Thus the casual effect being estimated is the average treatment effect of the treated.

We replicated the study conducted in \cite{bhagat_national_2017} by estimating balancing weights using two different methods. The first method uses entropy balancing (EB) where the estimated balancing weights are obtained with (\ref{dual-ebal}) and (\ref{weights-ebal}). Recall that these balancing weights applied to (\ref{HT}) is doubly-robust \citep{zhao_entropy_2017}. The second method fits a propensity score model using logistic regression (IPW). With the fitted propensity score, we then use (\ref{HT-ATT}) substituting $\hat{\pi}(\mathbf{X})$ for $\pi(\mathbf{X})$ to estimate $\tau_{\text{ATT}}$. The causal effect estimates using the propensity score matched (PSM) cohort from the original paper are also reported along with the unadjusted (UN) results in Table \ref{est-table}.

\begin{figure}[H]
	\centering
	\includegraphics[scale = 0.6]{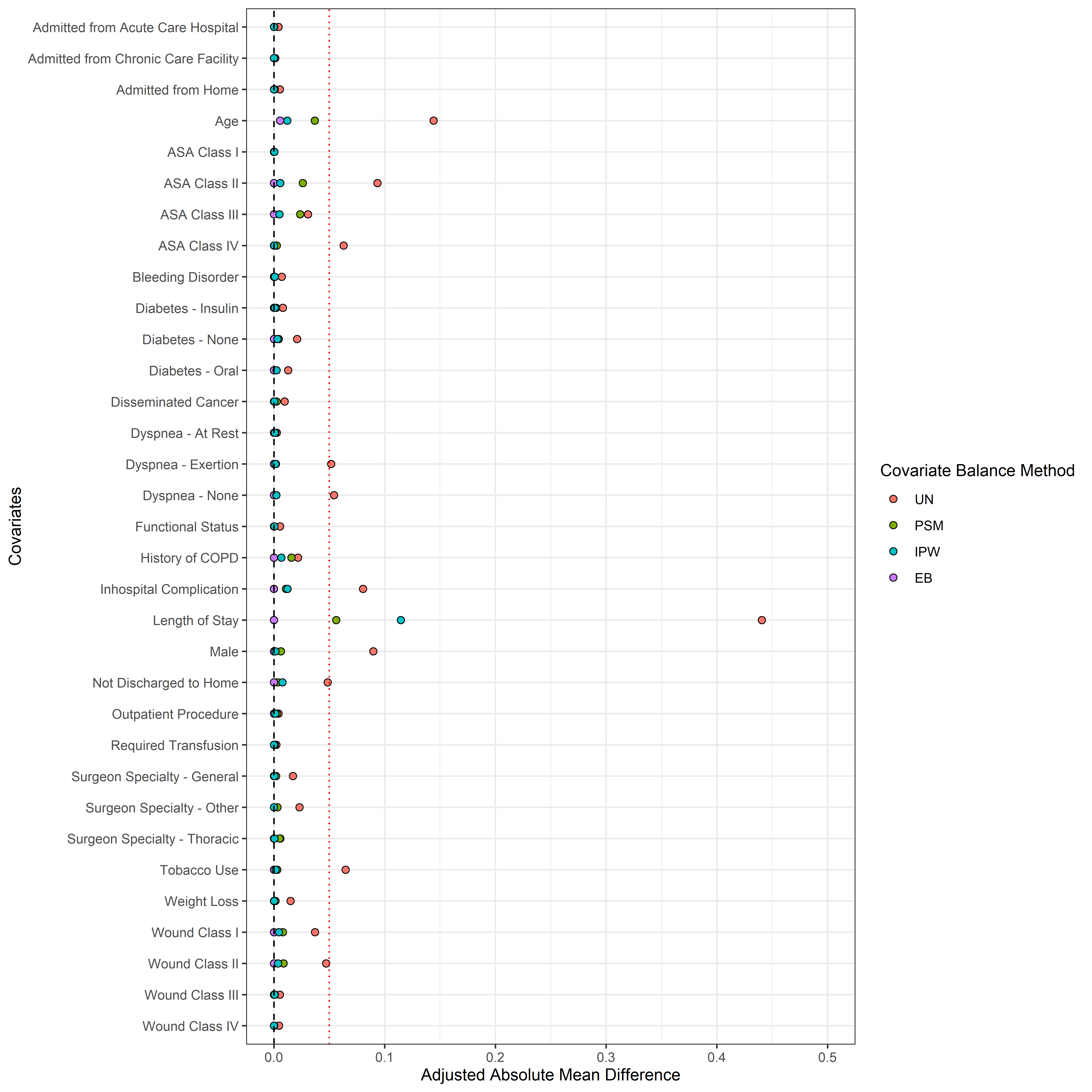}
	\caption{Each point represents the adjusted absolute standardized mean difference (x-axis) between thoracoscopic and open lung resection patients. The covariates in the plot (y-axis) are included into each model. The red dotted line marks an absolute standardized mean difference of 0.05.}\label{cobalt-plot}
\end{figure}

Figure \ref{cobalt-plot} shows the amount of imbalance observed for each of the covariates among those included in the covariate balancing models. We see that across each covariate, entropy balancing perfectly balances the first sample moments of the covariate distribution between the two treatment groups. Logistic regression appears to be less adequate at balancing the covariate moments than matching. However, aside from hospital length of stay, each of the weighted mean differences fell within the conservative 0.05 unit threshold using the inverse probability of treatment weights. The unadjusted differences do not share the same success as their adjusted counterparts, suggesting that some form of balancing should be implemented. After estimating $\tau_{\text{ATT}}$, notice in Table \ref{est-table} that the estimated risk difference is significant when using either the inverse probability of treatment weights or entropy balancing, but is not significant when using propensity score matching. The difference is likely due to the 2,712 patients that were omitted when matching. This discrepancy illuminates and emphasizes the importance of selecting the most appropriate method for balancing covariate data, even within a large observational study.

\section{Discussion}\label{discussion}

The generalized projection of a Bregman distance from a vector of sampling weights onto a set of intersecting hyperplanes is a powerful and flexible tool for normalizing data. In particular, this process is quite useful for constructing balancing weights for estimating causal effects. Using properties of dual optimization, we identify a doubly-robust estimator for $\tau_{\text{ATE}}$ and the optimally weighted average treatment effect \citep{crump_moving_2006} in Theorem \ref{dr-ate} and Corollary \ref{dr-owate}. We also show that the dual interpretation of improved CBPS \citep{fan_improving_2018} is doubly-robust in Theorem \ref{dr-hte}. In Theorem \ref{equiv-hte}, we present the conditions for which the balancing weights produced by (\ref{weights-ate}) are the same as the weights produced by (\ref{weights-calib}) suggested by \cite{chan_globally_2015}. When the true balance functions are unknown, we can use nonparametric methods similar to those suggested by \cite{hirano_efficient_2003} and \cite{chan_globally_2015} within our framework to achieve global efficiency.

In the simulation studies we conducted, we observed that the balancing weights that are typically used to estimate $\tau_{\text{OWATE}}$ had the best performance for estimating $\tau_{\text{ATE}}$ when Assumption \ref{constant} is satisfied. We also observed that including additional constraints as in (\ref{primal-ate}) sometimes had better performance than estimators that require fewer constraints, like CBPS. When we assume a linear conditional average treatment effect, our dual interpretation of iCBPS performed better than the analogous GMM estimator. We then apply our framework to a real data set of lung resection patients. Here we demonstrate how the choice of balancing method can have a critical impact on the results of a study.

There are several limitations to our proposed framework. First, each sampling unit's treatment assignment is assumed to be independent from the treatment assignment of the other sampling units. This assumption is sometimes called the no interference assumption. Health outcomes research is rich in observational data from the emergence of the electronic health record. While numerous in size, these datasets are more complex with patients being clustered within regions, hospitals, clinics, and/or practicing physicians. These are all factors that need to be accounted for in some way. How to extend these methods to clustered data settings is currently under investigation. Second, linear equality constraints are often quite stringent. If a particular covariate is difficult to balance, our proposed framework will sometimes fail to find the appropriate balancing weights. \cite{zubizarreta_stable_2015} proposes using stable balancing weights which places linear inequality constraints on the weighted sample moments of the covariate distribution while minimizing the Euclidean distance. In more recent work, \cite{wang_minimal_2019} have combined this interval constrained optimization approach with calibration estimators. There is also the issue where the balance functions that generate either the outcome or treatment assignment are high-dimensional. This problem is not examined in the presented work. \cite{ning_robust_2018} propose using an augmented approach with the covariate balance propensity scores of \cite{imai_covariate_2013} and \cite{fan_improving_2018} in the spirit of \cite{robins_estimation_1994} and \cite{farrell_robust_2015}. Their proposed methodology boasts compelling results as the dimension of covariate distribution increases. It is possible that our approach could be extended using some of the methods proposed by \cite{ning_robust_2018}.

In addition to addressing some of the limitations identified in the previous paragraph, in future work we would also like to expand these methods to incorporate multivalued treatment assignments. This would entail modifying the Horvitz-Thompson estimator and also requires extending the constraint matrix $\mathbf{A}$ and target margins $\mathbf{b}$ to facilitate covariate balance between all pairwise combinations of the treatment assignments. Finally, we would like to further investigate methods for generalizing causal effect estimates to a target population, which would involve estimating $\mathbf{q}$ prior to estimating $\mathbf{p}$.

\appendix

\section*{Acknowledgments}
\textbf{Special Thanks:} We would like to thank Dr. Robert Meguid and the Adult and Child Consortium for Health Outcomes Research and Delivery Science (ACCORDS) program for making the unplanned readmissions data available to us. We would also like to thank Dr. Peter DeWitt for his help in deploying the associated R package \texttt{cbal}. 

\vspace{1ex}

\noindent\textbf{Funding information:} This research is supported by a pilot grant from the Data Science to Patient Value (D2V) initiative from the University of Colorado. We acknowledge partial support from NSF award number DMS-1420451.

\vspace{1ex}

\noindent\textbf{Disclaimer:} This manuscript will be submitted to the Department of Biostatistics and Informatics in the Colorado School of Public Health, University of Colorado Denver, in partial fulfillment of the requirements for the degree of Doctor of Philosophy in Biostatistics for Kevin P. Josey.

\bibliographystyle{apalike}
\bibliography{simBib}

\section{R Package and Simulation Code}

The R package used to fit balancing weights as the generalized projection of Bregman distance is still in development with a working version available at \texttt{https://github.com/kevjosey/cbal}. The code used to conduct the simulation study in Section \ref{numerical} is available at the following URL: \texttt{https://github.com/kevjosey/cbal-sim}. The code for replicating the study by \cite{bhagat_national_2017} is available from the authors upon request.

\section{Technical Proofs}

\subsection{Proof of Lemma 1}\label{unique-proof}

\begin{proof}
For any $\mathbf{w} \in \Omega \cap \bar{\Delta}^n$, define
\[ \mathcal{S} \equiv \{\mathbf{p} \in \bar{\Delta}^n : \infdiv{\mathbf{p}}{\mathbf{q}} \le \infdiv{\mathbf{w}}{\mathbf{q}} \}. \]
The set $\mathcal{S}$ is bounded due to the requirement that Bregman distances have bounded partial level sets (Definition 2.1.1 of \cite{censor_parallel_1998}) and closed since $\infdiv{\mathbf{p}}{\mathbf{q}}$ is continuous in $\mathbf{p} \in \bar{\Delta}^n$. Therefore, the non-empty intersection $\mathcal{T} \equiv \Omega \cap \bar{\Delta}^n \cap \mathcal{S}$ is bounded. Since $\Omega \cap \bar{\Delta}^n$ and $\mathcal{S}$ are closed, $\mathcal{T}$ is also closed and hence compact. Thus, $\infdiv{\mathbf{p}}{\mathbf{q}}$, takes its infimum subject to $\mathbf{p} \in \Omega \cap \bar{\Delta}^n$ at $\hat{\mathbf{p}} \in \mathcal{T}$.

To prove that $\hat{\mathbf{p}} \in \Omega \cap \bar{\Delta}^n$ is unique, suppose there are two points $\hat{\mathbf{p}}, \tilde{\mathbf{p}} \in \Omega \cap \bar{\Delta}^n$  such that $\hat{\mathbf{p}} \ne \tilde{\mathbf{p}}$ and $\infdiv{\hat{\mathbf{p}}}{\mathbf{q}} = \infdiv{\tilde{\mathbf{p}}}{\mathbf{q}} = \argmin_{\mathbf{p} \in \Omega \cap \bar{\Delta}^n} \infdiv{\mathbf{p}}{\mathbf{q}}$, for some $\mathbf{q} \in \Delta^n$. By the convexity of $\Omega \cap \bar{\Delta}^n$, $(\hat{\mathbf{p}} + \tilde{\mathbf{p}})/2 \in \Omega \cap \bar{\Delta}^n$. By the strict convexity of $f$, we have
\[ \begin{split}
\infdivb{(\hat{\mathbf{p}} + \tilde{\mathbf{p}})/2}{\mathbf{q}} &= f\left[\frac{1}{2}(\hat{\mathbf{p}} + \tilde{\mathbf{p}})\right] - f(\mathbf{q}) - [\nabla f(\mathbf{q})]^{T}\left[\frac{1}{2}(\hat{\mathbf{p}} + \tilde{\mathbf{p}}) - \mathbf{q}\right] \\
&< \frac{1}{2}f(\hat{\mathbf{p}}) + \frac{1}{2}f(\tilde{\mathbf{p}}) - \frac{1}{2}f(\mathbf{q}) - \frac{1}{2}f(\mathbf{q}) \\
&\quad - \frac{1}{2}[\nabla f(\mathbf{q})]^{T}(\hat{\mathbf{p}} - \mathbf{q}) - \frac{1}{2}[\nabla f(\mathbf{q})]^{T}(\tilde{\mathbf{p}} - \mathbf{q}) \\
&= \frac{1}{2}\left[\infdiv{\hat{\mathbf{p}}}{\mathbf{q}} + \infdiv{\tilde{\mathbf{p}}}{\mathbf{q}}\right] = \min_{\mathbf{p} \in \Omega \cap \bar{\Delta}^n} \infdiv{\mathbf{p}}{\mathbf{q}},
\end{split} \]
which is a contradiction.
\end{proof}

\subsection{Proof of Proposition 1}\label{global-min-proof}

\begin{proof}
Assume $\hat{\mathbf{p}} = \argmin_{\mathbf{p} \in \Omega \cap \bar{\Delta}^n} \infdiv{\mathbf{p}}{\mathbf{q}}$ exists. By Lemma 1, $\hat{\mathbf{p}}$ is a global minimum subject to $\mathbf{p} \in \Omega$. Due to the zone consistency assumption, if $\mathbf{q} \in \Delta^n$ then the minimum must be contained in $\Delta^n$ and not $\bar{\Delta}^n$. This assumption allows us to apply the Lagrange multiplier theorem \citep{bertsekas_nonlinear_1999} which proves the existence of a unique $\hat{\boldsymbol{\lambda}} \in \Re^m$ that satisfies
\begin{equation}\label{first-order}
\nabla_{\mathbf{p}} \infdiv{\hat{\mathbf{p}}}{\mathbf{q}} + \mathbf{A}\hat{\boldsymbol{\lambda}} = \mathbf{0}_n. 
\end{equation} 
Note that due to the strict convexity of $f$, $\nabla f: \Delta^n \rightarrow \Re^n$ is strictly increasing and is therefore injective. Thus, the transformation $(\nabla f)^{-1}: \Re^n \rightarrow \Delta^n$ also has a unique mapping. If we solve (\ref{first-order}) for $\hat{\mathbf{p}}$, we get
\[ \hat{\mathbf{p}} = (\nabla f)^{-1}\left[\nabla f(\mathbf{q}) - \mathbf{A}\hat{\boldsymbol{\lambda}}\right] = P_f\left(\mathbf{q}, \mathbf{A}\hat{\boldsymbol{\lambda}}\right). \]
\end{proof}

\subsection{Proof of Proposition 2}\label{dual-proof}

\begin{proof}
Let $\boldsymbol{\lambda} \in \Re^m$ and $\hat{\mathbf{p}} = \argmin_{\mathbf{p} \in \bar{\Delta}^n} L(\mathbf{p}, \mathbf{q}, \boldsymbol{\lambda})$. Since $f$ is zone consistent, a minimum of the Lagrangian with respect to $\mathbf{p} \in \Delta^n$ must satisfy
\[ \nabla f(\hat{\mathbf{p}}) = \nabla f(\mathbf{q}) - \mathbf{A}\boldsymbol{\lambda}. \]
By the strict convexity of $f$ (see the proof in Section \ref{global-min-proof}), we can solve for $\hat{\mathbf{p}}$ to get
\begin{equation}\label{lagrange-min}
\hat{\mathbf{p}} = (\nabla f)^{-1} [\nabla f(\mathbf{q}) - \mathbf{A}\boldsymbol{\lambda}] = \argmin_{\mathbf{p} \in \Delta^n} L(\mathbf{p}, \mathbf{q}, \boldsymbol{\lambda}).
\end{equation}

From (\ref{lagrange-min}) we know that for all $\mathbf{p} \in \Omega \cap \Delta^n$, \[ L\left[P_f(\mathbf{q}, \mathbf{A}\boldsymbol{\lambda}), \mathbf{q}, \boldsymbol{\lambda}\right] \le \infdiv{\mathbf{p}}{\mathbf{q}} + \left(\mathbf{A}^{T}\mathbf{p} - \mathbf{b}\right)^{T}\boldsymbol{\lambda} = \infdiv{\mathbf{p}}{\mathbf{q}}. \] According to the definition of the generalized projection, minimizing the right hand side subject to $\mathbf{p} \in \Omega \cap \Delta^n$ yields the solution to the primal problem and
\begin{equation}\label{inequality}
L\left[P_f(\mathbf{q}, \mathbf{A}\boldsymbol{\lambda}), \mathbf{q}, \boldsymbol{\lambda}\right] \le \infdiv{\hat{\mathbf{p}}}{\mathbf{q}}
\end{equation}
for all $\boldsymbol{\lambda} \in \Re^m$. By Proposition 1, there exists some $\hat{\boldsymbol{\lambda}} \in \Re^m$ such that $\hat{\mathbf{p}} = P_f\left(\mathbf{q}, \mathbf{A}\hat{\boldsymbol{\lambda}}\right)$. Therefore,
\begin{equation}\label{equivalence}
\begin{split}
\infdiv{\hat{\mathbf{p}}}{\mathbf{q}}
&= \infdiv{\hat{\mathbf{p}}}{\mathbf{q}} + \left(\mathbf{A}^{T}\hat{\mathbf{p}} - \mathbf{b}\right)^{T}\hat{\boldsymbol{\lambda}} \\
&= L(\hat{\mathbf{p}}, \mathbf{q}, \hat{\boldsymbol{\lambda}}) \\
&= L\left[P_f\left(\mathbf{q}, \mathbf{A}\hat{\boldsymbol{\lambda}}\right), \mathbf{q}, \hat{\boldsymbol{\lambda}}\right].
\end{split}
\end{equation}
Substituting the result of (\ref{equivalence}) into (\ref{inequality}), we have 
\[ L\left[P_f(\mathbf{q}, \mathbf{A}\boldsymbol{\lambda}), \mathbf{q}, \boldsymbol{\lambda}\right] \le L\left[P_f\left(\mathbf{q}, \mathbf{A}\hat{\boldsymbol{\lambda}}\right), \mathbf{q}, \hat{\boldsymbol{\lambda}}\right] \]
for all $\boldsymbol{\lambda} \in \Re^m$ which implies
\[ \hat{\boldsymbol{\lambda}} = \argmax_{\boldsymbol{\lambda} \in \Re^m} L\left[P_f(\mathbf{q}, \mathbf{A}\boldsymbol{\lambda}), \boldsymbol{\lambda}\right]. \]

\end{proof}

\subsection{Proof of Theorem 1}\label{dr-ate-proof}

\begin{proof}
Denote $\tau^*= \mathbb{E}[Y(1) - Y(0)]$ and suppose
\begin{equation}\label{out_1}
\begin{split}
\mu^*_0(\mathbf{X}_i) &= \sum_{j = 1}^{m} c_j(\mathbf{X}_i)\beta^*_j  \enskip \text{and} \\
\mu^*_1(\mathbf{X}_i) &= \tau^* + \sum_{j = 1}^{m} c_j(\mathbf{X}_i)\beta^*_j 
\end{split}
\end{equation}
where $\beta^*_j \in \Re$, $j = 1,2,\ldots,m$, denotes the true coefficient values of the outcome model. Let $\hat{p}(\mathbf{X}_i)$ be determined by (\ref{weights-cbps}) where $\hat{\boldsymbol{\lambda}}$ is solved with (\ref{dual-cbps}).

Recall that another doubly-robust estimator for the average treatment effect is \[ \hat{\tau}_{\text{AIPW}} = \frac{1}{n}\sum_{i = 1}^n \left\{ \frac{Z_iY_i}{\hat{\pi}(\mathbf{X}_i)} - \frac{[Z_i - \hat{\pi}(\mathbf{X}_i)]\hat{\mu}_1(\mathbf{X}_i)}{\hat{\pi}(\mathbf{X}_i)}  - \frac{(1 - Z_i)Y_i}{1 - \hat{\pi}(\mathbf{X}_i)} - \frac{[Z_i - \hat{\pi}(\mathbf{X}_i)]\hat{\mu}_0(\mathbf{X}_i)}{1 - \hat{\pi}(\mathbf{X}_i)}\right\} \] where 
\begin{equation}\label{out_est}
\begin{split}
\hat{\mu}_0(\mathbf{X}_i) &= \sum_{j = 1}^{m} c_j(\mathbf{X}_i)\hat{\beta}_j  \enskip \text{and} \\
\hat{\mu}_1(\mathbf{X}_i) &= \hat{\tau}_{\text{G}} + \sum_{j = 1}^{m} c_j(\mathbf{X}_i)\hat{\beta}_j 
\end{split}
\end{equation}
so that $\hat{\mu}_0(\mathbf{X}_i) \rightarrow_p \mu^*_0(\mathbf{X}_i)$ and $\hat{\mu}_1(\mathbf{X}_i) \rightarrow_p \mu^*_1(\mathbf{X}_i)$. If we substitute $\hat{p}(\mathbf{X}_i)$ for $\hat{\pi}(\mathbf{X}_i)^{-1}$ and $[1 - \hat{\pi}(\mathbf{X}_i)]^{-1}$ and assume, without loss of generality, that $\sum_{i = 1}^n Z_i\hat{p}(\mathbf{X}_i) = n$, we get
\[ \begin{split}
\hat{\tau} - \hat{\tau}_{\text{AIPW}}
&= \frac{1}{n} \sum_{i = 1}^n \hat{p}(\mathbf{X}_i) Z_i \hat{\mu}_1(\mathbf{X}_i) - \frac{1}{n}\sum_{i = 1}^n \hat{p}(\mathbf{X}_i) (1 - Z_i) \hat{\mu}_0(\mathbf{X}_i) - \hat{\tau}_{\text{G}} \\
&= \frac{1}{n}\sum_{i = 1}^n \hat{p}(\mathbf{X}_i) Z_i \hat{\tau}_{\text{G}} + \frac{1}{n}\sum_{j = 1}^m \hat{\beta}_j \left[\sum_{i = 1}^n \hat{p}(\mathbf{X}_i)(2Z_i - 1)c_j(\mathbf{X}_i)\right] - \hat{\tau}_{\text{G}} \\
&= \frac{1}{n}\sum_{i = 1}^n \hat{p}(\mathbf{X}_i) Z_i \hat{\tau}_{\text{G}} - \hat{\tau}_{\text{G}} = 0. \end{split} \]
This shows that the balancing weight method is algebraically equivalent to the augmented regression approach of \cite{robins_estimation_1994} under the condition that the outcome model is correctly specified. A similar proof was applied to entropy balancing for estimating $\tau_{\text{ATT}}$ in \cite{zhao_entropy_2017}.

Now, let $\text{logit}[\pi^*(\mathbf{X})] = \text{logit}[\pi(\mathbf{X}; \boldsymbol{\lambda}^*)] = \sum_{j = 1}^m c_j(\mathbf{X})\lambda^*_j$ and suppose $\mu_1(\mathbf{X})$ and $\mu_0(\mathbf{X})$ are unknown. Denote the $(m \times 1)$ vector of the linearly independent balance functions as \[ \mathbf{c}(\mathbf{X}) \equiv [c_1(\mathbf{X}), c_2(\mathbf{X}), \ldots, c_m(\mathbf{X})]^T. \] To show consistency under these conditions, we follow the approach of \cite{tsiatis_semiparametric_2006} and \cite{kennedy_semiparametric_2016} by employing influence functions about the parameters. Define the estimating equation for $\boldsymbol{\lambda}$ as 
\begin{equation}\label{zeta}
\begin{split} \boldsymbol{\zeta}(\mathbf{X}, Z; \boldsymbol{\lambda}) 
&\equiv Z\left\{1 + \exp\left[ -\sum_{j = 1}^m c_j(\mathbf{X})\lambda_j\right]\right\}\mathbf{c}(\mathbf{X}) \\ 
&\quad - (1-Z)\left\{1 + \exp\left[ \sum_{j = 1}^m  c_j(\mathbf{X})\lambda_j\right]\right\}\mathbf{c}(\mathbf{X}),
\end{split}
\end{equation}
which is the first order condition of (\ref{dual-cbps}). As a result of Propositions 1 and 2, $\hat{\boldsymbol{\lambda}}$ must satisfy \[ \sum_{i = 1}^n \boldsymbol{\zeta}(\mathbf{X}_i, Z_i; \hat{\boldsymbol{\lambda}}) = \mathbf{0}_m. \]
Next, define the estimating equation for $\tau$ as
\begin{equation}\label{psi}
\begin{split}
\psi(\mathbf{X}, Y, Z; \boldsymbol{\lambda}, \tau) 
&\equiv Z\left\{1 + \exp\left[ -\sum_{j = 1}^m c_j(\mathbf{X})\lambda_j\right]\right\} \left[Y(1) - \tau\right] \\
&\quad - (1 - Z)\left\{1 + \exp\left[ \sum_{j = 1}^m c_j(\mathbf{X})\lambda_j\right]\right\} Y(0).
\end{split}
\end{equation}
The estimating equation defined in (\ref{psi}) is solved with the Horvitz-Thompson estimator which satisfies
\[ \sum_{i = 1}^n \psi(\mathbf{X}_i, Y_i, Z_i; \hat{\boldsymbol{\lambda}}, \hat{\tau}) = 0. \]
Define $\boldsymbol{\theta} \equiv (\boldsymbol{\lambda}^{T}, \tau)^{T}$ and the the stacked estimating equations as \[ \boldsymbol{\xi}(\mathbf{X}, Z, Y; \boldsymbol{\theta}) \equiv [\boldsymbol{\zeta}(\mathbf{X}, Z; \boldsymbol{\lambda})^{T}, \psi(\mathbf{X}, Y, Z; \boldsymbol{\lambda}, \tau)]^{T}.\] Under standard regularity assumptions described in \cite{tsiatis_semiparametric_2006}, it can be shown using Taylor's theorem that
\begin{equation}\label{influence} \hat{\boldsymbol{\theta}} - \boldsymbol{\theta}^* = -\left\{\mathbb{E}\left[\frac{\partial \boldsymbol{\xi}(\mathbf{X}, Z, Y; \boldsymbol{\theta}^*)}{\partial \boldsymbol{\theta}}\right]\right\}^{-1} \left\{\sum_{i = 1}^n \boldsymbol{\xi}(\mathbf{X}_i, Z_i, Y_i; \boldsymbol{\theta}^*)\right\} + o_p\left(n^{-1/2}\right).
\end{equation}
It is then straightforward to identify the influence function for $\tau$ from (\ref{influence}), which we denote as
\begin{equation}\label{phi}
\begin{split}
\phi(\mathbf{X}, Y, Z; \boldsymbol{\lambda}^*, \tau^*) &\equiv \psi(\mathbf{X}, Y, Z; \boldsymbol{\lambda}^*, \tau^*)  \\
&\quad - \mathbb{E}\left[\frac{\partial\psi(\mathbf{X}, Y, Z; \boldsymbol{\lambda}^*, \tau^*)}{\partial \boldsymbol{\lambda}}\right]^{T} \left\{\mathbb{E}\left[\frac{\partial \boldsymbol{\zeta}(\mathbf{X}, Z; \boldsymbol{\lambda}^*)}{\partial \boldsymbol{\lambda}}\right]\right\}^{-1} \boldsymbol{\zeta}(\mathbf{X}, Z; \boldsymbol{\lambda}^*). 
\end{split}
\end{equation}
This allows us to express
\begin{equation}\label{tau-inf}
\begin{split}
\hat{\tau}-\tau^* &= \frac{1}{n}\sum_{i = 1}^n \psi(\mathbf{X}_i, Y_i, Z_i; \boldsymbol{\lambda}^*, \tau^*)  + o_p\left(n^{-1/2}\right) \\
&\quad - \mathbb{E}\left[\frac{\partial\psi(\mathbf{X}, Y, Z; \boldsymbol{\lambda}^*, \tau^*)}{\partial \boldsymbol{\lambda}}\right]^{T} \left\{\mathbb{E}\left[\frac{\partial\boldsymbol{\zeta}(\mathbf{X}, Z; \boldsymbol{\lambda}^*)}{\partial \boldsymbol{\lambda}}\right]\right\}^{-1}\left[\frac{1}{n}\sum_{i = 1}^n \boldsymbol{\zeta}(\mathbf{X}_i, Z_i; \boldsymbol{\lambda}^*) \right].
\end{split}
\end{equation}
When $\text{logit}[\pi^*(\mathbf{X})] = \sum_{j = 1}^m c_j(\mathbf{X})\lambda^*_j$, then $\mathbb{E}\left[\boldsymbol{\zeta}(\mathbf{X}, Z; \boldsymbol{\lambda}^*)\right] = \mathbf{0}_m$. In a similar fashion, and given Assumption 2, it is trivial to show $\mathbb{E}\left[ \psi(\mathbf{X}, Y, Z; \boldsymbol{\lambda}^*, \tau^*) \right] = 0$. After applying the weak law of large numbers to (\ref{tau-inf}), we conclude $\hat{\tau} \rightarrow_p \tau^*$.  Notice that Assumption 3 is not required to prove this result.
 
For simplicity, denote $\phi^*(\mathbf{X}, Y, Z) = \phi(\mathbf{X}, Y, Z; \boldsymbol{\lambda}^*, \boldsymbol{\tau}^*)$. Following the classical central limit theorem, the influence function for $\tau$ can also be used to derive the asymptotic distribution which is \[ \sqrt{n}(\hat{\tau} - \tau^*) \rightarrow_d \mathcal{N}\left\{0, \mathbb{E}[\phi^*(\mathbf{X}, Y, Z)^2]\right\}. \] To simplify the algebra further, we introduce a few nuisance parameters and rewrite (\ref{zeta}) and (\ref{psi}). Define $\boldsymbol{\gamma} \equiv \mathbb{E}[\mathbf{c}(\mathbf{X})]$ and $\mu_0 \equiv \mathbb{E}[Y(0)] = \mathbb{E}[\mu_0(\mathbf{X})]$. If $\mathbf{c}(\mathbf{X})$ includes an intercept term (i.e. $c_1(\mathbf{X}) = 1$), then by Propositions 1 and 2 we can equivalently express (\ref{zeta}) as 
\[ \begin{split}
\boldsymbol{\zeta}(\mathbf{X}, Z; \boldsymbol{\lambda}) 
&\equiv Z\left\{1 + \exp\left[ -\sum_{j = 1}^m c_j(\mathbf{X})\lambda_j\right]\right\} \left[\mathbf{c}(\mathbf{X}) - \boldsymbol{\gamma} \right] \\ 
&\quad - (1-Z)\left\{1 + \exp\left[ \sum_{j = 1}^m  c_j(\mathbf{X})\lambda_j\right]\right\}\left[\mathbf{c}(\mathbf{X}) - \boldsymbol{\gamma} \right]
\end{split} \]
for any $\boldsymbol{\gamma} \in \Re^m$. Equation (\ref{psi}) can be rewritten as
\[ \begin{split} \psi(\mathbf{X}, Y, Z; \boldsymbol{\lambda}, \tau) 
&\equiv Z\left\{1 + \exp\left[ -\sum_{j = 1}^m c_j(\mathbf{X})\lambda_j\right]\right\} \left[Y(1) - \mu_0 - \tau \right]\\
&\quad - (1 - Z)\left\{1 + \exp\left[ \sum_{j = 1}^m c_j(\mathbf{X})\lambda_j\right]\right\} \left[Y(0) - \mu_0 \right] \end{split} \]
for any $\mu_0 \in \Re$. Moreover, let $\mathbf{H}_{\mathbf{U}, \mathbf{V}}(W) = \mathbb{E}\{W[\mathbf{U} - \mathbb{E}(\mathbf{U})][\mathbf{V} - \mathbb{E}(\mathbf{V})]^{T}\}$, $\mathbf{H}_{\mathbf{U}}(W) = \mathbf{H}_{\mathbf{U}, \mathbf{U}}(W)$ and \[ \mathbf{M} = \mathbf{H}_{\mathbf{c}(\mathbf{X}), \mu_0(\mathbf{X})}[\pi^*(\mathbf{X})] + \mathbf{H}_{\mathbf{c}(\mathbf{X}), \mu_1(\mathbf{X})}[1 - \pi^*(\mathbf{X})]. \] Given Assumption 2, the asymptotic variance is expanded into
\begin{equation}\label{gnarly-variance}
\begin{split}
\mathbb{E}\left[\phi^*(\mathbf{X}, Y, Z)^2\right] &= \mathbf{M}^{T}\mathbf{H}^{-1}_{\mathbf{c}(\mathbf{X})}(1) \bigg(\mathbf{H}_{\mathbf{c}(\mathbf{X})}\left\{\frac{1}{\pi^*(\mathbf{X})[1 - \pi^*(\mathbf{X})]}\right\} \mathbf{H}^{-1}_{\mathbf{c}(\mathbf{X})}(1) \mathbf{M} \\
&\quad\quad - 2\mathbf{H}_{\mathbf{c}(\mathbf{X}), \mu_0(\mathbf{X})}\left[\frac{1}{1 - \pi^*(\mathbf{X})}\right] - 2\mathbf{H}_{\mathbf{c}(\mathbf{X}), \mu_1(\mathbf{X})}\left[\frac{1}{\pi^*(\mathbf{X})}\right] \bigg) \\
&\ + H_{Y(0)}\left[\frac{1}{1 - \pi^*(\mathbf{X})}\right] + H_{Y(1)}\left[\frac{1}{\pi^*(\mathbf{X})}\right].
\end{split}
\end{equation}
If we suppose $\mu_1(\mathbf{X}) = \mu^*_1(\mathbf{X})$ and $\mu_0(\mathbf{X}) = \mu^*_0(\mathbf{X})$  as in (\ref{out_1}), we can write \[ \mathbf{H}_{\mathbf{c}(\mathbf{X}),\mu^*_0(\mathbf{X})}(W) = \mathbf{H}_{\mathbf{c}(\mathbf{X}), \mu^*_1(\mathbf{X})}(W) = \mathbf{H}_{\mathbf{c}(\mathbf{X})}(W)\boldsymbol{\beta}^*. \] This allows us to reduce (\ref{gnarly-variance}) into the more succinct format of
\[ \begin{split}
\mathbb{E}\left[\phi^*(\mathbf{X}, Y, Z)^2\right] 
&= H_{Y(0)}\left[\frac{1}{1 - \pi^*(\mathbf{X})}\right] + H_{Y(1)}\left[\frac{1}{\pi^*(\mathbf{X})}\right] - H_{\mu^*_0(\mathbf{X})}\left\{\frac{1}{\pi^*(\mathbf{X})[1 - \pi^*(\mathbf{X})]}\right\} \\
&= \mathbb{E}\left\{\frac{\mathbb{V}[Y(1)|\mathbf{X}]}{\pi^*(\mathbf{X})} + \frac{\mathbb{V}[Y(0)|\mathbf{X}]}{1 - \pi^*(\mathbf{X})}\right\}
\end{split} \]
where the second equality holds due to the conditional variance identity.
\end{proof}

\begin{remark}
We avoid using sampling weights in Theorem 1 due to the observation that \[ \frac{1}{1 + (q-1)\exp\left[-\sum_{j = 1}^n c_j(\mathbf{X}) \lambda_{j}\right]} \ne 1 - \frac{1}{1 + (q-1)\exp\left[\sum_{j = 1}^n c_j(\mathbf{X})\lambda_{j}\right]} \] for some $q \in (1,\infty)$. To resolve this issue, we define $q(1)$ and $q(0)$ such that $q(1) - 1 = [q(0)- 1]^{-1}$, which can be used to identify the sampling weights as $q \equiv Zq(1) + (1 - Z)q(0)$. It is straightforward to show \[ \frac{1}{1 + [q(1)-1]\exp\left[-\sum_{j = 1}^n c_j(\mathbf{X}) \lambda_{j}\right]} = 1 - \frac{1}{1 + [q(0)-1]\exp\left[\sum_{j = 1}^n c_j(\mathbf{X})\lambda_{j}\right]}. \] We would then assume $\pi^*(\mathbf{X})^{-1} = 1 + [q(1)-1]\exp\left[-\sum_{j = 1}^n c_j(\mathbf{X})\lambda_{j}\right]$ in condition 1 of Theorem 1 and substitute this notation accordingly into the proof.
\end{remark}

\subsection{Proof of Corollary 1}\label{dr-owate-proof}

\begin{proof}
Since the constraints placed on the weighting estimator in Theorem 1 are identical to the constraints in Corollary 1, the consistency of the Horvitz-Thompson estimator under Assumption 3 and the condition that $\mu_0(\mathbf{X})$ is correctly specified is proven in a nearly identical manner to Theorem 1 using balancing weights that are instead determined by (\ref{weights-owate}) and (\ref{dual-owate}). The only difference would be to use an augmented regression estimator for $\tau_{\text{OWATE}}$ instead of $\tau_{\text{ATE}}$.

Assume $\text{logit}[\pi^*(\mathbf{X})] = \text{logit}[\pi(\mathbf{X}; \boldsymbol{\lambda}^*)] = \sum_{j = 1}^m c_j(\mathbf{X})\lambda^*_j$ and let \[ \tau^*  = \frac{\mathbb{E}\{\pi^*(\mathbf{X}) [1 - \pi^*(\mathbf{X}) ] [Y(1) - Y(0)]\}}{ \mathbb{E}\{\pi^*(\mathbf{X}) [1 - \pi^*(\mathbf{X})]\}} . \] Our approach to proving consistency under the condition that the treatment assignment is correctly specified is similar to the proof in Section \ref{dr-ate-proof}. However, we instead define
\[ \boldsymbol{\zeta}'(\mathbf{X}, Z; \boldsymbol{\lambda})  \equiv \frac{\left\{\frac{Z\mathbf{c}(\mathbf{X})}{1 + \exp\left[\sum_{j = 1}^m c_j(\mathbf{X})\lambda_j\right]} - \frac{(1 - Z)\mathbf{c}(\mathbf{X})}{1 + \exp\left[-\sum_{j = 1}^m c_j(\mathbf{X})\lambda_j\right]} \right\}}{\mathbb{E}\left\{\frac{1}{2 + \exp\left[-\sum_{j = 1}^m c_j(\mathbf{X})\lambda_j\right] + \exp\left[\sum_{j = 1}^m c_j(\mathbf{X})\lambda_j\right]} \right\}}, \]
which is the first order condition of (\ref{dual-owate}) including a normalizing constant in the denominator \citep{hirano_efficient_2003}, and  \[ \psi'(\mathbf{X}, Y, Z; \boldsymbol{\lambda}, \tau) \equiv \frac{\left\{\frac{Z\left[Y(1) - \tau\right]}{1 + \exp\left[\sum_{j = 1}^m c_j(\mathbf{X})\lambda_j\right]} - \frac{(1 - Z)Y(0)}{1 + \exp\left[-\sum_{j = 1}^m c_j(\mathbf{X})\lambda_j\right]} \right\}}{{\mathbb{E}\left\{\frac{1}{2 + \exp\left[-\sum_{j = 1}^m c_j(\mathbf{X})\lambda_j\right] + \exp\left[\sum_{j = 1}^m c_j(\mathbf{X})\lambda_j\right]} \right\}}}.\] From the proof of Theorem 1, the influence function for $\tau$ is identified as
\[ \begin{split}
\phi'(\mathbf{X}, Y, Z; \boldsymbol{\lambda}^*, \tau^*) &\equiv \psi'(\mathbf{X}, Y, Z; \boldsymbol{\lambda}^*, \tau^* ) \\
&\quad - \mathbb{E}\left[\frac{\partial\psi'(\mathbf{X}, Y, Z; \boldsymbol{\lambda}^*, \tau^*)}{\partial \boldsymbol{\lambda}}\right]^{T} \left\{\mathbb{E}\left[\frac{\partial \boldsymbol{\zeta}'(\mathbf{X}, Z; \boldsymbol{\lambda}^*)}{\partial \boldsymbol{\lambda}}\right]\right\}^{-1} \boldsymbol{\zeta}'(\mathbf{X}, Z; \boldsymbol{\lambda}^*).
\end{split} \]
When $\text{logit}[\pi^*(\mathbf{X})] = \sum_{j = 1}^m c_j(\mathbf{X})\lambda^*_j$ and given Assumption 2, then
\[ \begin{split} 
\mathbb{E}\left[\psi'(\mathbf{X}, Y, Z; \boldsymbol{\lambda}^*, \tau^* )\right] 
&= \frac{\mathbb{E} \left\{ \pi^*(\mathbf{X})[1 - \pi^*(\mathbf{X})] \left[\mu_1(\mathbf{X}) - \mu_0(\mathbf{X}) - \tau^*\right]\right\}}{\mathbb{E}\{\pi^*(\mathbf{X}) [1 - \pi^*(\mathbf{X})]\}} \\
&=  \frac{\mathbb{E} \left\{ \pi^*(\mathbf{X})[1 - \pi^*(\mathbf{X})] \left[\mu_1(\mathbf{X}) - \mu_0(\mathbf{X}) - \mu_1(\mathbf{X}) + \mu_0(\mathbf{X}) \right]\right\}}{\mathbb{E}\{\pi^*(\mathbf{X}) [1 - \pi^*(\mathbf{X})]\}} 
\end{split}\] 
so that $\mathbb{E}\left[\psi'(\mathbf{X}, Y, Z; \boldsymbol{\lambda}^*, \tau^* )\right] = 0$. In a similar fashion, we can show $\mathbb{E}\left[ \boldsymbol{\zeta}'(\mathbf{X}, Z; \boldsymbol{\lambda}^*) \right] = \mathbf{0}_m$. We then apply the weak law of large numbers to conclude $\hat{\tau} \rightarrow_p \tau^*$.  

\begin{remark}
If we set $\tau^* = \mathbb{E}[Y(1) - Y(0)] = \mathbb{E}[\mu_1(\mathbf{X}) - \mu_0(\mathbf{X})]$, then we would require Assumption 3 to hold in order for $\mathbb{E}\left[\psi'(\mathbf{X}, Y, Z; \boldsymbol{\lambda}^*, \tau^*)\right] = 0$.
\end{remark}
 
Recall from the proof of Theorem 1 that
\[ \sqrt{n}(\hat{\tau} - \tau^*) \rightarrow_d \mathcal{N}\left\{0, \mathbb{E}\left[\phi'(\mathbf{X}, Y, Z; \boldsymbol{\lambda}^*, \tau^* )^2\right]\right\}.\]
Notice that \[ \psi'(\mathbf{X}, Y, Z; \boldsymbol{\lambda}^*, \tau^* ) = \frac{\pi^*(\mathbf{X})[1 - \pi^*(\mathbf{X})]\psi(\mathbf{X}, Y, Z; \boldsymbol{\lambda}^*, \tau^* )}{\mathbb{E}\{\pi^*(\mathbf{X}) [1 - \pi^*(\mathbf{X})]\}}\] where $\psi(\mathbf{X}, Y, Z; \boldsymbol{\lambda}, \tau)$ is defined in (\ref{psi}) and \[ \boldsymbol{\zeta}'(\mathbf{X}, Z; \boldsymbol{\lambda}^*) = \frac{\pi^*(\mathbf{X})[1 - \pi^*(\mathbf{X})]\boldsymbol{\zeta}(\mathbf{X}, Z; \boldsymbol{\lambda}^*)}{\mathbb{E}\{\pi^*(\mathbf{X}) [1 - \pi^*(\mathbf{X})]\}}\] where $\boldsymbol{\zeta}(\mathbf{X}, Z; \boldsymbol{\lambda})$ is defined in (\ref{zeta}). If $\mu^*_1(\mathbf{X})$ and $\mu^*_0(\mathbf{X})$ follow the form of (\ref{out_1}), then
\[ \mathbb{E}\left[\frac{\partial\psi'(\mathbf{X}, Y, Z; \boldsymbol{\lambda}^*, \tau^*)}{\partial \boldsymbol{\lambda}}\right]^{T} \left\{\mathbb{E}\left[\frac{\partial \boldsymbol{\zeta}'(\mathbf{X}, Z; \boldsymbol{\lambda}^*)}{\partial \boldsymbol{\lambda}}\right]\right\}^{-1} = (\boldsymbol{\beta}^*)^{T}. \]
This means \[ \phi'(\mathbf{X}, Y, Z; \boldsymbol{\lambda}^*, \tau^* ) = \frac{[\pi^*(\mathbf{X})(1 - \pi^*(\mathbf{X})]}{\mathbb{E}\{\pi^*(\mathbf{X}) [1 - \pi^*(\mathbf{X})]\}}\phi(\mathbf{X}, Y, Z; \boldsymbol{\lambda}^*, \tau^*)\] where $\phi(\mathbf{X}, Y, Z; \boldsymbol{\lambda}, \tau)$ is defined in (\ref{phi}). According to the proof of Theorem 1, we know \[ \mathbb{E}\left[\phi(\mathbf{X}, Y, Z; \boldsymbol{\lambda}^*, \tau^* )^2 \middle|\mathbf{X}\right] = \frac{\mathbb{V}[Y(1)|\mathbf{X}]}{\pi^*(\mathbf{X})} + \frac{\mathbb{V}[Y(0)|\mathbf{X}]}{1 - \pi^*(\mathbf{X})} \] which allows us to write
\[\begin{split}
\mathbb{E}\left[\phi'(\mathbf{X}, Y, Z; \boldsymbol{\lambda}^*, \tau^* )^2\right] 
&= \frac{\mathbb{E}\left\{\pi^*(\mathbf{X})^2[1 - \pi^*(\mathbf{X})]^2 \phi(\mathbf{X}, Y, Z; \boldsymbol{\lambda}^*, \tau^* )^2\right\}}{\mathbb{E}\{\pi^*(\mathbf{X}) [1 - \pi^*(\mathbf{X})]\}^2} \\
&= \frac{\mathbb{E}\left(\left\{\pi(\mathbf{X})[1 - \pi(\mathbf{X})]\right\}^2 \left\{\frac{\mathbb{V}[Y(1)|\mathbf{X}]}{\pi(\mathbf{X})} + \frac{\mathbb{V}[Y(0)|\mathbf{X}]}{1 - \pi(\mathbf{X})}\right\}\right)}{\mathbb{E}\{\pi^*(\mathbf{X}) [1 - \pi^*(\mathbf{X})]\}^2}
\end{split} \]
\end{proof}

\begin{remark}
The proof of Theorem 1, and by extension Corollary 1, exploits the identity
\begin{equation}\label{ps-identity}
P_f\left[q, \sum_{j = 1}^n c_j(\mathbf{X})\lambda^*_j\right]\pi^*(\mathbf{X}) = P_f\left[q, -\sum_{j = 1}^n c_j(\mathbf{X})\lambda^*_j\right][1 - \pi^*(\mathbf{X})] = \omega^*(\mathbf{X}),
\end{equation}
where $\omega^*(\mathbf{X})$ is a known weighting function. In the case of Theorem 1, $\omega^*(\mathbf{X}) = 1$, while for Corollary 1, $\omega^*(\mathbf{X}) = \pi^*(\mathbf{X})[1 - \pi^*(\mathbf{X})]$. In the case where we find estimators for $\tau_{\text{ATT}}$, we require \[P_f\left[q, \sum_{j = 1}^n c_j(\mathbf{X})\lambda^*_j\right][1 - \pi^*(\mathbf{X})] = \pi^*(\mathbf{X}).\] The identity in (\ref{ps-identity}) requires careful selection of the generalized projection, which should complement the assumed functional form of the propensity score. To generalize Theorem 1 would require modifying the statement of condition 1 to assume $\Pr\{Z = 1|\mathbf{X}\} = P_f\left[q, \sum_{j = 1}^n c_j(\mathbf{X})\lambda_{j}\right]^{-1}$ and $\Pr\{Z = 0|\mathbf{X}\} = P_f\left[q, -\sum_{j = 1}^n c_j(\mathbf{X})\lambda_{j}\right]^{-1}$. If this condition is satisfactory, the proof of Theorem 1 would be the same albeit with changes to the notation. However, we found this condition to be counter-intuitive in instances where we violate (\ref{ps-identity}).
\end{remark}

\subsection{Proof of Theorem 2}\label{dr-hte-proof}

\begin{proof}
First, we will prove that $\hat{\tau}$ is consistent for $\tau_{\text{ATE}}$ when $\mu_0(\mathbf{X})$ lies on the span of a linearly independent set of balance functions, $\{c_j(\mathbf{X}) : j = 1,2,\ldots,m\}$. Let $\hat{p}(\mathbf{X}_i)$ be determined by (\ref{weights-icbps}) where $\hat{\boldsymbol{\lambda}}$ is solved by (\ref{dual-icbps}). Assume that the conditional means of the potential outcomes are
\begin{equation}\label{out_2}
\begin{split}
\mu^*_0(\mathbf{X}_i) &= \sum_{j = 1}^{m} c_j(\mathbf{X}_i)\beta^*_j \enskip \text{and} \\
\mu^*_1(\mathbf{X}_i) &= \sum_{j = 1}^{m} c_j(\mathbf{X}_i)\alpha^*_j  + \sum_{j = 1}^{m} c_j(\mathbf{X}_i)\beta^*_j 
\end{split}
\end{equation}
where $\alpha^*_j$ and $\beta^*_j$ are the true values of the regression coefficients, and $\tau^* = \mathbb{E}\left[\mu^*_1(\mathbf{X}) - \mu^*_0(\mathbf{X})\right]$. Notice that if $c_1(\mathbf{X}_i) = 1$ for all $i = 1, 2,\ldots,n$, then $n = \sum_{i = 1}^n \hat{p}(\mathbf{X}_i)Z_i$. We will again show the algebraic equivalency of the exact balancing weight estimator with the doubly-robust estimator \[ \hat{\tau}_{\text{AIPW}} = \frac{1}{n} \sum_{i = 1}^n \left\{ \frac{Z_iY_i}{\hat{\pi}(\mathbf{X}_i)} - \frac{[Z_i - \hat{\pi}(\mathbf{X}_i)]\hat{\mu}_1(\mathbf{X}_i)}{\hat{\pi}(\mathbf{X}_i)}  - \frac{(1 - Z_i)Y_i}{1 - \hat{\pi}(\mathbf{X}_i)} - \frac{[Z_i - \hat{\pi}(\mathbf{X}_i)]\hat{\mu}_0(\mathbf{X}_i)}{1 - \hat{\pi}(\mathbf{X}_i)}\right\} \] where
\[ \begin{split}
\hat{\mu}_0(\mathbf{X}_i) &= \sum_{j = 1}^{m} c_j(\mathbf{X}_i)\hat{\beta}_j \enskip \text{and} \\
\hat{\mu}_1(\mathbf{X}_i) &= \sum_{j = 1}^{m} c_j(\mathbf{X}_i)\left( \hat{\alpha}_j + \hat{\beta}_j \right)
\end{split} \] so that $\hat{\mu}_0(\mathbf{X}_i) \rightarrow_p \mu^*_0(\mathbf{X}_i)$ and $\hat{\mu}_1(\mathbf{X}_i) \rightarrow_p \mu^*_1(\mathbf{X}_i)$. We then substitute $\hat{p}(\mathbf{X}_i)$ for $\hat{\pi}(\mathbf{X}_i)^{-1}$ and $[1 - \hat{\pi}(\mathbf{X}_i)]^{-1}$ depending to whether $Z_i = 1$ and $Z_i = 0$, respectively. This gives us
\[ \begin{split}
\hat{\tau} - \hat{\tau}_{\text{AIPW}}
&= \frac{1}{n}\sum_{i = 1}^n \hat{p}(\mathbf{X}_i) Z_i \hat{\mu}_1(\mathbf{X}_i) - \frac{1}{n}\sum_{i = 1}^n \hat{p}(\mathbf{X}_i) (1 - Z_i) \hat{\mu}_0(\mathbf{X}_i) - \frac{1}{n}\sum_{i = 1}^n \sum_{j = 1}^m c_j(\mathbf{X}_i)\hat{\alpha}_j \\
&= \frac{1}{n} \sum_{j = 1}^m \hat{\beta}_j \left[\sum_{i = 1}^n \hat{p}(\mathbf{X}_i)(2Z_i - 1)c_j(\mathbf{X}_i)\right] + \frac{1}{n} \sum_{j = 1}^m \hat{\alpha}_j \left[\sum_{i = 1}^n\hat{p}(\mathbf{X}_i)Z_ic_j(\mathbf{X}_i) - c_j(\mathbf{X}_i)\right] \\
&= 0. \end{split} \]

Now we let $\text{logit}[\pi^*(\mathbf{X})] = \text{logit}[\pi(\mathbf{X}; \boldsymbol{\lambda}^*_0)] = \sum_{j = 1}^m c_j(\mathbf{X})\lambda^*_{j0}$ and assume $\mu_1(\mathbf{X})$ and $\mu_0(\mathbf{X})$ are unknown. We will reuse much of the notation and definitions presented in the proof of Theorem 1. First, we redefine the estimating equations for $\boldsymbol{\lambda}$ as
\[ \begin{split}
\boldsymbol{\zeta}_0(\mathbf{X}, Z; \boldsymbol{\lambda}) 
&\equiv Z\left\{1 + \exp\left[ -\sum_{j = 1}^m c_j(\mathbf{X})(\lambda_{j0} + \lambda_{j1})\right]\right\}\mathbf{c}(\mathbf{X}) \\
&\quad - (1-Z)\left\{1 + \exp\left[\sum_{j = 1}^m  c_j(\mathbf{X})\lambda_{j0}\right]\right\}\mathbf{c}(\mathbf{X}) \enskip \text{and} \\
\boldsymbol{\zeta}_1(\mathbf{X}, Z; \boldsymbol{\lambda})
&\equiv Z\left\{1 + \exp\left[ -\sum_{j = 1}^m c_j(\mathbf{X})(\lambda_{j0} + \lambda_{j1})\right]\right\} \mathbf{c}(\mathbf{X}) - \mathbf{c}(\mathbf{X})
\end{split} \]
with $\boldsymbol{\zeta}(\mathbf{X}, Z; \boldsymbol{\lambda}) = \left[\boldsymbol{\zeta}_0(\mathbf{X}, Z; \boldsymbol{\lambda})^{T}, \boldsymbol{\zeta}_1(\mathbf{X}, Z; \boldsymbol{\lambda})^{T}\right]^{T}$. We also redefine the estimating equation for $\tau$ as
\[ \begin{split} 
\psi(\mathbf{X}, Y, Z; \boldsymbol{\lambda}, \tau) 
&\equiv Z\left\{1 + \exp\left[ -\sum_{j = 1}^m c_j(\mathbf{X})(\lambda_{j0} + \lambda_{j1})\right]\right\} \left[Y(1) - \tau\right] \\
&\quad - (1 - Z)\left\{1 + \exp\left[ \sum_{j = 1}^m c_j(\mathbf{X})\lambda_{j0}\right]\right\} Y(0). 
\end{split} \]
Recall from the proof of Theorem 1 that under the standard regularity assumptions of \cite{tsiatis_semiparametric_2006}, we have
\begin{equation}\label{tau-inf-2}
\begin{split}
\hat{\tau} - \tau^* &= \frac{1}{n}\sum_{i = 1}^n \psi(\mathbf{X}_i, Y_i, Z_i; \boldsymbol{\lambda}^*, \tau^*)  + o_p\left(n^{-1/2}\right) \\
&\quad - \mathbb{E}\left[\frac{\partial\psi(\mathbf{X}, Y, Z; \boldsymbol{\lambda}^*, \tau^*)}{\partial \boldsymbol{\lambda}}\right]^{T} \left\{\mathbb{E}\left[\frac{\partial\boldsymbol{\zeta}(\mathbf{X}, Z; \boldsymbol{\lambda}^*)}{\partial \boldsymbol{\lambda}}\right]\right\}^{-1}\left[\frac{1}{n}\sum_{i = 1}^n \boldsymbol{\zeta}(\mathbf{X}_i, Z_i; \boldsymbol{\lambda}^*), \right].
\end{split}
\end{equation}
Next, observe that
\begin{equation}\label{annoying}
\mathbb{E}\left[\boldsymbol{\zeta}_1(\mathbf{X}, Z; \boldsymbol{\lambda}^*)\right] = \mathbb{E}\left\{\frac{1 + \exp\left[-\sum_{j = 1}^m c_j(\mathbf{X})(\lambda^*_{j0} + \lambda^*_{j1})\right]}{1 + \exp\left[-\sum_{j = 1}^m c_j(\mathbf{X})\lambda^*_{j0}\right]} \mathbf{c}(\mathbf{X}) - \mathbf{c}(\mathbf{X})\right\}.
\end{equation} 
The previously mentioned regularity assumptions allow us to interchange differentiation and expectation to obtain the first derivative of (\ref{annoying}) with respect to $\boldsymbol{\lambda}_1$, which yields a negative-definite matrix (given that the balance functions are linearly independent). This implies (\ref{annoying}) can only be zero if $\lambda^*_{j1} = 0$ for all $j = 1,2,\ldots,m$. It is then trivial to show $\mathbb{E}[\boldsymbol{\zeta}_0(\mathbf{X}, Z; \boldsymbol{\lambda}^*)] = 0$ and $\mathbb{E}[\psi(\mathbf{X}, Y, Z; \boldsymbol{\lambda}^*, \tau^*)] = 0$. This means that the expected value of (\ref{tau-inf-2}) is zero, and we can apply the weak law of large numbers to arrive at our desired result where $\hat{\tau} \rightarrow_p \tau^*$.

Due to the central limit theorem, we know $\sqrt{n}(\hat{\tau} - \tau^*) \rightarrow_d \mathcal{N}\left\{0, \mathbb{E}\left[\phi^*(\mathbf{X}, Y, Z)^2\right]\right\}$ where $\phi^*(\mathbf{X}, Y, Z) = \phi(\mathbf{X}, Y, Z; \boldsymbol{\lambda}^*, \tau^*)$ is defined in (\ref{phi}). Without loss of generality (and by the assumption that $c_1(\mathbf{X}) = 1$) redefine the estimating equations
\[ \begin{split}
\boldsymbol{\zeta}_0(\mathbf{X}, Z; \boldsymbol{\lambda}) 
&\equiv Z\left\{1 + \exp\left[ -\sum_{j = 1}^m c_j(\mathbf{X})(\lambda_{j0} + \lambda_{j1})\right]\right\}[\mathbf{c}(\mathbf{X}) - \boldsymbol{\gamma}] \\
&\quad - (1-Z)\left\{1 + \exp\left[ \sum_{j = 1}^m  c_j(\mathbf{X})\lambda_{j0}\right]\right\} [\mathbf{c}(\mathbf{X}) - \boldsymbol{\gamma}], \\
\boldsymbol{\zeta}_1(\mathbf{X}, Z; \boldsymbol{\lambda}) 
&\equiv Z\left\{1 + \exp\left[ -\sum_{j = 1}^m c_j(\mathbf{X})(\lambda_{j0} + \lambda_{j1})\right]\right\}[\mathbf{c}(\mathbf{X}) - \boldsymbol{\gamma}] - [\mathbf{c}(\mathbf{X}) - \boldsymbol{\gamma}]
\end{split} \]
for some $\boldsymbol{\gamma} \in \Re^{m}$, and
\[ \begin{split}
\psi(\mathbf{X}, Y, Z; \boldsymbol{\lambda}, \tau) 
&\equiv Z\left\{1 + \exp\left[ -\sum_{j = 1}^m c_j(\mathbf{X})(\lambda_{j0} + \lambda_{j1})\right]\right\} \left[Y(1) - \mu_0 - \tau\right] \\
&\quad - (1 - Z)\left\{1 + \exp\left[ \sum_{j = 1}^m c_j(\mathbf{X})\lambda_{j0}\right]\right\} [Y(0) - \mu_0]. 
\end{split} \]
for some $\mu_0 \in \Re$. To lighten the notation further, we will write $\psi^* \equiv \psi(\mathbf{X}, Y, Z; \boldsymbol{\lambda}^*, \tau^*)$ and $\boldsymbol{\zeta}^* \equiv \boldsymbol{\zeta}(\mathbf{X}, Z; \boldsymbol{\lambda}^*) $. We then expand the asymptotic variance into more digestible components with
\begin{equation}\label{gnarly-variance-2}
\begin{split}
\mathbb{E}\left[\phi^*(\mathbf{X}, Y, Z)^2\right] &= \mathbb{E}\left[(\boldsymbol{\psi}^*)^2\right] - 2\mathbb{E}\left(\frac{\partial \psi^*}{\partial \boldsymbol{\lambda}}\right)^{T}\left[\mathbb{E}\left(\frac{\partial \boldsymbol{\zeta}^*}{\partial \boldsymbol{\lambda}}\right)\right]^{-1}\mathbb{E}\left(\boldsymbol{\zeta}^*\psi^*\right) \\
&\quad +\mathbb{E}\left(\frac{\partial \psi^*}{\partial \boldsymbol{\lambda}}\right)^{T}\left[\mathbb{E}\left(\frac{\partial \boldsymbol{\zeta}^*}{\partial \boldsymbol{\lambda}}\right)\right]^{-1}\mathbb{E}\left[\boldsymbol{\zeta}^*(\boldsymbol{\zeta}^*)^{T}\right]\left[\mathbb{E}\left(\frac{\partial \boldsymbol{\zeta}^*}{\partial \boldsymbol{\lambda}}\right)\right]^{-1}\mathbb{E}\left(\frac{\partial \psi^*}{\partial \boldsymbol{\lambda}}\right)
\end{split}
\end{equation}
Suppose $\mu_1(\mathbf{X}) = \mu^*_1(\mathbf{X})$ and $\mu_0(\mathbf{X}) = \mu^*_0(\mathbf{X})$ which are defined in (\ref{out_2}). Then \[ \mathbb{E}\left(\frac{\partial \psi^*}{\partial \boldsymbol{\lambda}}\right)^{T}\left[\mathbb{E}\left(\frac{\partial \boldsymbol{\zeta}^*}{\partial \boldsymbol{\lambda}}\right)\right]^{-1} = \left[(\boldsymbol{\beta}^{*})^T, (\boldsymbol{\alpha}^{*})^{T}\right] \] and, with some algebra,
\[ \begin{split} 
\mathbb{E}\left(\boldsymbol{\zeta}^*\psi^*\right) &= \mathbb{E}\left[\boldsymbol{\zeta}^*(\boldsymbol{\zeta}^*)^{T}\right]\left[\mathbb{E}\left(\frac{\partial \boldsymbol{\zeta}^*}{\partial \boldsymbol{\lambda}}\right)\right]^{-1}\mathbb{E}\left(\frac{\partial \psi^*}{\partial \boldsymbol{\lambda}}\right) \\
&= \begin{bmatrix}
\mathbf{H}_{\mathbf{c}(\mathbf{X})}\left[\frac{1}{\pi^*(\mathbf{X})}\right]\boldsymbol{\alpha}^* + \mathbf{H}_{\mathbf{c}(\mathbf{X})}\left\{\frac{1}{\pi^*(\mathbf{X})[ 1 - \pi^*(\mathbf{X})]}\right\}\boldsymbol{\beta}^* \\
\mathbf{H}_{\mathbf{c}(\mathbf{X})}\left[\frac{1 - \pi^*(\mathbf{X})}{\pi^*(\mathbf{X})}\right]\boldsymbol{\alpha}^* + \mathbf{H}_{\mathbf{c}(\mathbf{X})}\left[\frac{1}{\pi^*(\mathbf{X})}\right]\boldsymbol{\beta}^*
\end{bmatrix}.
\end{split}\]
This allows us to further reduce (\ref{gnarly-variance-2}) into  
\[\begin{split} 
\mathbb{E}\left[\phi^*(\mathbf{X}, Y, Z)^2\right]
&= H_{Y(1)}\left[\frac{1}{\pi^*(\mathbf{X})}\right] + H_{Y(0)}\left[\frac{1}{1 - \pi^*(\mathbf{X})}\right] \\
&\quad  - \begin{bmatrix} (\boldsymbol{\beta}^{*})^{T}  & (\boldsymbol{\alpha}^{*})^{T}  \end{bmatrix}\begin{bmatrix}
\mathbf{H}_{\mathbf{c}(\mathbf{X})}\left[\frac{1}{\pi^*(\mathbf{X})}\right]\boldsymbol{\alpha}^* + \mathbf{H}_{\mathbf{c}(\mathbf{X})}\left\{\frac{1}{\pi^*(\mathbf{X})[ 1 - \pi^*(\mathbf{X})]}\right\}\boldsymbol{\beta}^* \\
\mathbf{H}_{\mathbf{c}(\mathbf{X})}\left[\frac{1 - \pi^*(\mathbf{X})}{\pi^*(\mathbf{X})}\right]\boldsymbol{\alpha}^* + \mathbf{H}_{\mathbf{c}(\mathbf{X})}\left[\frac{1}{\pi^*(\mathbf{X})}\right]\boldsymbol{\beta}^*
\end{bmatrix} \\
&= H_{Y(1)}\left[\frac{1}{\pi^*(\mathbf{X})}\right] + H_{Y(0)}\left[\frac{1}{1 - \pi^*(\mathbf{X})}\right] - H_{\mu^*_1(\mathbf{X})}\left[\frac{1}{\pi^*(\mathbf{X})}\right] \\
&\quad - H_{\mu^*_0(\mathbf{X})}\left[\frac{1}{1 - \pi^*(\mathbf{X})}\right] + (\boldsymbol{\alpha}^{*})^{T}\mathbf{H}_{\mathbf{c}(\mathbf{X})}\left(1\right)\boldsymbol{\alpha}^* \\
&= \mathbb{E}\left\{\frac{\mathbb{V}[Y(1)|\mathbf{X}]}{\pi^*(\mathbf{X})} + \frac{\mathbb{V}[Y(0)|\mathbf{X}]}{1 - \pi^*(\mathbf{X})} + \left(\sum_{j = 1}^m c_j(\mathbf{X})\alpha^*_j - \tau^* \right)^2\right\} 
\end{split}\]
where the last equality holds due to the conditional variance identity and by recognizing \[ \mathbb{E}\left[\left(\sum_{j = 1}^m c_j(\mathbf{X})\alpha^*_j - \tau \right)^2\right] =  (\boldsymbol{\alpha}^{*})^{T}\mathbf{H}_{\mathbf{c}(\mathbf{X})}\left(1\right)\boldsymbol{\alpha}^* \]
\end{proof}

\subsection{Proof of Theorem 3}

\begin{proof}
Let $\mathbf{p} \in \Omega_1$ which satisfies the constraints 
\[ \begin{split} 
\sum_{i = 1}^n p_iZ_ic_j(\mathbf{X}_i) &= b_j \enskip \text{and} \\
\sum_{i = 1}^n p_iZ_ic_j(\mathbf{X}_i) &= \sum_{i = 1}^n p_i(1 - Z_i)c_j(\mathbf{X}_i).
\end{split} \] 
This implies $\sum_{i = 1}^n p_i(1 - Z_i)c_j(\mathbf{X}_i) = b_j$ so that $\mathbf{p} \in \Omega_0$. Thus $\Omega_1 = \Omega_0$. 

According to Lemma 1, the generalized projection is unique implying \[ \hat{\mathbf{p}} = \argmin_{\mathbf{p} \in \Omega_1 \cap \Delta^n} \infdiv{\mathbf{p}}{\mathbf{q}} = \argmin_{\mathbf{p} \in \Omega_0 \cap \Delta^n} \infdiv{\mathbf{p}}{\mathbf{q}}. \] Because $G$ is a monotone increasing transformation of $D_f$, the optimal point is preserved and we conclude
\[ \hat{\mathbf{p}} = \argmin_{\mathbf{p} \in \Omega_1 \cap \Delta^n} \infdiv{\mathbf{p}}{\mathbf{q}} = \argmin_{\mathbf{p} \in \Omega_0 \cap \Delta^n} G(\mathbf{p}) = \tilde{\mathbf{p}}. \]
\end{proof}

\newpage
\section*{Tables and Figures}

\begin{table}[H]
\scriptsize
\centering
\begin{tabular}{ccccccc}
\hline
\multicolumn{1}{l}{} & \begin{tabular}[c]{@{}c@{}}Outcome\\ Scenario\end{tabular} & \begin{tabular}[c]{@{}c@{}}Treatment\\ Assignment\\ Scenario\end{tabular} & IPW & CBPS & SENT &  BENT \\ \hline
\multirow{4}{*}{\begin{tabular}[c]{@{}c@{}}Avg. Estimate \\ (MC Std. Err.)\end{tabular}} 
 & a & a & 19.60 (3.28) & 20.02 (0.55) & 20.02 (0.53) & 20.02 (0.51) \\
 & a & b & 20.40 (2.70) & 20.00 (0.50) & 20.01 (0.49) & 20.00 (0.49) \\
 & b & a & 19.36 (5.27) & 19.67 (3.81) & 19.75 (2.90) & 19.90 (2.21) \\
 & b & b & 15.33 (3.57) & 15.26 (3.28) & 15.83 (2.73) & 16.47 (2.31) \\
 &  &  &  &  &  &  \\
\multirow{4}{*}{\begin{tabular}[c]{@{}c@{}}Mean Squared\\ Error (Bias)\end{tabular}} 
 & a & a & 10.91 (-0.40) & 0.30 (0.02) & 0.28 (0.02) & 0.26 (0.02) \\
 & a & b & 7.42 (0.40) & 0.25 (0.00) & 0.24 (0.01) &  0.24 (0.00) \\
 & b & a & 28.21 (-0.64) & 14.63 (-0.33) & 8.49 (-0.25) & 4.87 (-0.10) \\
 & b & b & 34.54 (-4.67) & 33.14 (-4.74) & 24.79 (-4.17) & 17.83 (-3.53) \\ \hline
\end{tabular}
\caption{Average estimate, Monte Carlo standard error, residual mean squared error, and empirical bias of the constant conditional ATE using the four methods for estimating balancing weights described in Section \ref{homogeneous}. IPW uses inverse probability of treatment weights estimated from a generalized linear model, CBPS uses the covariate balance propensity score weights, SENT uses the constrained optimal solution of the shifted relative entropy, and BENT uses the constrained optimal solution of the binary relative entropy.}\label{sim-table-1}
\end{table}

\newpage
\begin{table}[H]
\scriptsize
\centering
\begin{tabular}{cccccccc}
\hline
\multicolumn{1}{l}{} & \begin{tabular}[c]{@{}c@{}}Outcome\\ Scenario\end{tabular} & \begin{tabular}[c]{@{}c@{}}Treatment\\ Assignment\\ Scenario\end{tabular} & AIPW & CAL & iCBPS & hdCBPS & SENT \\ \hline
\multirow{4}{*}{\begin{tabular}[c]{@{}c@{}}Avg. Estimate \\ (MC Std. Err.)\end{tabular}} 
 & a & a & 19.93 (1.51) & 19.93 (1.49) & 19.06 (3.94) & 19.93 (1.51) & 19.93 (1.49) \\
 & a & b & 20.09 (1.48) & 20.09 (1.47) & 18.05 (3.50) & 20.09 (1.47) & 20.09 (1.47) \\
 & b & a & 19.80 (7.35) & 20.43 (2.76) & 19.72 (3.63) & 19.59 (3.02) & 20.25 (2.78) \\
 & b & b & 15.04 (4.52) & 16.91 (2.55) & 14.21 (4.48) & 16.15 (2.76) & 16.77 (2.54) \\
 &  &  &  &  &  &  \\
\multirow{4}{*}{\begin{tabular}[c]{@{}c@{}}Mean Squared\\ Error (Bias)\end{tabular}} 
 & a & a & 2.28 (-0.07) & 2.22 (-0.07) & 16.41 (-0.94) & 2.28 (-0.07) & 2.23 (-0.07) \\
 & a & b & 2.18 (0.09) & 2.16 (0.09) & 16.08 (-1.95) & 2.17 (0.09) & 2.16 (0.09) \\
 & b & a & 53.98 (-0.20) & 7.77 (0.43) & 13.26 (-0.28) & 9.28 (-0.41) & 7.78 (0.25) \\
 & b & b & 45.04 (-4.96) & 16.01 (-3.09) & 53.58 (-5.79) & 22.45 (-3.85) & 16.84 (-3.23) \\ \hline
\end{tabular}
\caption{Average estimate, Monte Carlo standard error, residual mean squared error, and empirical bias of the linear conditional ATE using the five methods for estimating balancing weights described in Section \ref{heterogeneous}. AIPW uses an augmented inverse probability of treatment approach where the propensity scores are estimated from a generalized linear model, CAL uses the calibration estimated weights, iCBPS uses the covariate balance propensity score weights, hdCBPS is an augmented version of CBPS, and SENT uses the constrained optimal solution of the shifted relative entropy.}\label{sim-table-2}
\end{table}

\newpage
\begin{table}[H]
\centering
\begin{tabular}{cccc}
\hline
Balancing Method & Risk Difference & Std. Error & 95\% Confidence Interval \\ \hline
UN & -0.021 & 0.006 & (-0.032, -0.010) \\
PSM & -0.010 & 0.006 & (-0.021, 0.001) \\
IPW & -0.015 & 0.006 & (-0.027, -0.004) \\
EB & -0.016 & 0.006 & (-0.027, -0.004) \\ \hline
\end{tabular}
\caption{ATT estimates for 30-day unplanned readmission in thoracoscopic versus open lung resection patients. UN denotes the unadjusted results, PSM denotes the propensity score matched results, IPW denotes the inverse probability weighted results, EB denotes the entropy balancing results.}\label{est-table}
\end{table}

\newpage

\newpage

\newpage

\newpage

\end{document}